\documentclass[final,3p,twocolumn]{elsarticle}

\usepackage{amsmath}
\usepackage{txfonts}
\usepackage[utf8]{inputenc}
\usepackage{graphicx}

\usepackage{hyperref,xcolor}
\usepackage[T1]{fontenc} 
\usepackage{graphicx}
\usepackage{ulem}

\usepackage{amssymb} 
\usepackage{tabularx} 
\usepackage{gensymb}
\usepackage{appendix}
\usepackage{enumerate}   
\usepackage{bm}
\usepackage{setspace}
\usepackage{textcomp}

\usepackage{subcaption}
\usepackage{caption}

\journal{Astroparticle Physics}

\begin{document}

\begin{frontmatter}

\title{In-ice Radio Signatures of Cosmic Ray Particle Cascades
\\ }

\author[a]{Simon Chiche}
\author[a]{Simona Toscano}
\author[b]{Krijn D. de Vries}

\address[a]{Inter-University Institute For High Energies (IIHE), Universit\'{e} libre de Bruxelles (ULB),
Boulevard du Triomphe 2, 1050 Brussels, Belgium}
\address[b]{Vrije Universiteit Brussel, Physics Department, Pleinlaan 2, 1050 Brussels, Belgium}

\date{Received $\langle$date$\rangle$ / Accepted $\langle$date$\rangle$}

\begin{abstract}

To detect ultra-high-energy neutrinos, experiments such as the Askaryan Radio Array and the Radio Neutrino Observatory in Greenland target the radio emission induced by these particles as they cascade in the ice. This is done by, amongst others, using deep in-ice antennas at the South Pole or in Greenland.  A crucial step toward this goal is the characterization of the in-ice radio emission from cosmic-ray–induced particle showers. These showers form a primary background for neutrino searches, but can also be used to validate the detection principle and provide calibration signals for in-ice radio detectors. In this work, we use the  Monte-Carlo framework FAERIE to perform the first characterization of cosmic ray signals with  simulations that incorporate both their in-air and in-ice emissions. We investigate cosmic ray signatures such as their radiation energy, timing, polarization, and frequency spectrum and quantify how they depend on shower properties. These results provide key guidelines for cosmic-ray identification and cosmic-ray/neutrino discrimination in future in-ice radio experiments.

\end{abstract}

\begin{keyword}
Radio-detection, In-ice experiments, High-energy astroparticles, Monte-Carlo simulations
\vspace{0.5cm}
\noindent

\end{keyword}

\end{frontmatter}


\newpage

\newcommand{\red}{\color{red}}


\section{Introduction}~\label{sec:Intro}

Ultra-high energy neutrinos ($E>10^{16}\, \rm eV$) are unique messengers for probing the high energy universe. These particles travel with no deflection nor attenuation and serve as an unambiguous  probe of hadronic acceleration~\cite{UHENU2022JHEAp..36...55A}. The first detection of an ultra-high energy neutrino by KM3NeT in 2023~\cite{KM3NeT:2025npi} has provided a proof of their existence, further motivating the need for their detection. Yet, neutrino detection is challenging and requires to build experiments with unprecedented sensitivity to address the low fluxes at the highest energies and open a new astronomical window~\cite{Guepin2022NatRP...4..697G}. In-ice radio detection of astroparticles is  a particularly promising technique. When a neutrino interacts in the ice, it induces a cascade of secondary particles that emit radio waves through the Askaryan mechanism~\cite{Askaryan:1961pfb, JaimeAskaryan2011PhRvD..84j3003A,Coleman2025APh...17203136C}, coherent up to around one GHz. These radio waves can then be detected with radio antennas buried deep inside the ice sheet. The attenuation length of radio waves in polar ice reaches the kilometer scale, typically an order of magnitude larger than that of optical light, which is limited to about 100 m~\cite{OpticalAttenuation2006JGRD..11113203A}. This kilometer-scale transparency makes radio detection particularly well suited for instrumenting very large effective volumes at relatively low cost~\cite{AttenuationBarwick:2005zz}. Current experiments such as the Askaryan Radio Array~\cite{ARA2016PhRvD..93h2003A} (ARA) and next-generation  detectors including the Radio Neutrino Observatory in Greenland~\cite{RNOG2021JInst..16P3025A} (RNO-G), and IceCube-Gen2 radio~\cite{ICGen22021JPhG...48f0501A,Gen2Radio2025arXiv250707813G}, therefore aim to deploy radio antennas in Greenland or at the South Pole to detect ultra-high energy neutrinos. Such a detection is not trivial as it requires to identify neutrinos solely from their radio signatures despite the numerous sources of noise in the radio domain. Cosmic rays in particular are a major background since they can produce radio waves that resemble those induced by neutrino in-ice interactions and that can also reach the deep antennas. The cosmic ray flux is expected to be at least 1000 times larger than the neutrino flux which implies that cosmic ray/neutrino discrimination is essential to detect ultra-high energy neutrinos. On the other hand, cosmic ray detection would provide a unique opportunity to validate the detection principle of in-ice experiments, as recently reported by the ARA collaboration~\cite{ARAcr2025arXiv251021104A}, and help calibrate the detectors. FAERIE~\cite{FAERIE2024PhRvD.110b3010D} is a numerical tool that combines both CORSIKA7~\cite{Corsika1998cmcc.book.....H} and GEANT4~\cite{GEANT42016NIMPA.835..186A} Monte-Carlo codes, coupled with the CoREAS radio extension~\cite{COREAS2013AIPC.1535..128H}, to allow us to simulate the radio emission from cosmic ray particle cascades, as seen by deep in-ice observers, as predicted in~\cite{KDV2016APh....74...96D,Kockere1.2022PhRvD.106d3023D, FAERIE2024PhRvD.110b3010D}. In this work, we use FAERIE to perform an exhaustive characterization of the radio emission from cosmic-ray-induced particle cascades. Section~\ref{sec:Charac} presents the principles of cosmic-ray radio emission. In Section~\ref{sec:FAERIE}, we describe the construction of the cosmic-ray simulation library with FAERIE. Sections~\ref{sec:Amplitude} and~\ref{sec:RadE} explore the main properties of the radio signal, including its spatial amplitude distribution and radiation energy. Section~\ref{sec:Observables} specifically investigates the frequency spectrum, polarization, and timing of the signal, and identifies signatures for cosmic-ray detection and cosmic-ray/neutrino discrimination with in-ice radio detectors. Our results are discussed in Section~\ref{sec:Discussion}, before concluding in Section~\ref{sec:Conclusion}.



\section{Emission principle}~\label{sec:Charac}

\begin{figure}[tb]  
    \centering
    \includegraphics[width=\linewidth]{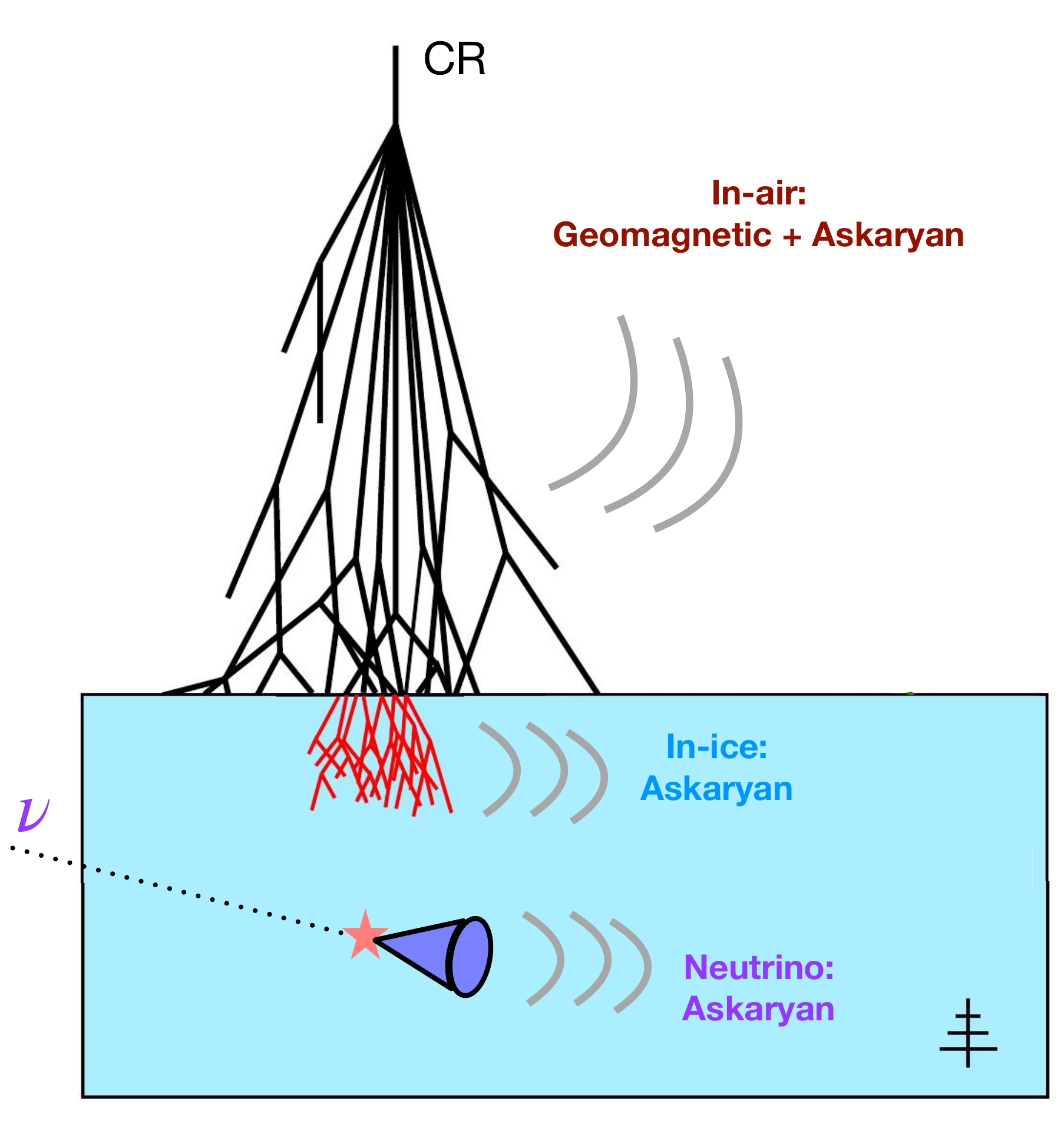}
    \caption{Sketch of a cosmic ray particle cascade and associated emissions. A first emission comes from the in-air cascade  (black lines). This emission propagates in the atmosphere and is transmitted to the ice. Another emission is generated directly by the in-ice cascade (red lines) and can also reach deep in-ice antennas. For comparison a typical neutrino event is also shown.}
    \label{fig:CRsketch}
\end{figure}
\subsection{Cosmic ray particle cascades}~\label{sec:Emission}
We illustrate the principles of cosmic ray radio emission in Fig.~\ref{fig:CRsketch}. When a cosmic ray enters the atmosphere, it induces an air shower, i.e., a cascade of secondary particles~\cite{Schroeder2017PrPNP..93....1S}. This cascade emits radio waves~\cite{HuegeDigital2016PhR...620....1H} that can propagate from the air to the ice and reach deep in-ice antennas, as indicated by the red ``in-air'' emission on the sketch. Additionally, if the shower is energetic enough, some particles from the air shower can also penetrate the ice and trigger an in-ice particle cascade~\cite{Kockere1.2022PhRvD.106d3023D}, which is likely to happen due to the high altitude above sea level of the ice sheet in polar areas. This secondary cascade will also emit radio waves, as indicated by the blue ``in-ice'' emission on the sketch. 

\subsection{In-air emission}

The radio emission from the in-air cascade is typical of an air shower and is the result of two main mechanisms: the geomagnetic and the Askaryan emissions~\cite{GeoScholten2008APh....29...94S, HuegeDigital2016PhR...620....1H}. The geomagnetic emission is the dominant mechanism. It originates from the deflection of the lightest charged particles in the shower, namely electrons and positrons, by the Earth’s magnetic field. This deflection creates a transverse current orthogonal to the shower propagation axis. The time variation of this current results in radio emission linearly polarized in the $\mathbf{v} \times \mathbf{B}$ direction, where $\mathbf{v}$ is the direction of propagation of the shower and $\mathbf{B}$ corresponds to the local direction of the Earth's magnetic field. The Askaryan emission~\cite{Askaryan:1961pfb, JaimeAskaryan2011PhRvD..84j3003A} is subdominant in air, it comes from the ionization of the propagation medium by high-energy shower particles which creates a net negative charge building up at the shower front. The time variation of this negative charge yields a radio emission radially polarized in a plane perpendicular to the shower propagation axis. 

\subsection{In-ice emission}

For the radio emission from the in-ice cascade, the Askaryan mechanism is the only contribution, as the geomagnetic mechanism is suppressed in dense media, since the shower length scale is much smaller, thus reducing the corresponding transverse current. The total (broadband) Askaryan power emitted in ice is also expected to be significantly larger than that of the in-air Askaryan component, since the emission scales with the negative charge excess at the shower front, which itself increases with the density of the medium in which the shower develops~\cite{Askaryan2011PhRvD..84j3003A, GlaserRadE2016JCAP...09..024G, PolarChiche2022APh...13902696C}. We note, however, that this scaling does not necessarily hold within a given frequency band, as a higher density reduces the typical shower track lengths and modifies the coherence conditions, thereby shifting the frequencies at which the emission is coherent~\cite{JuanDensity2023JCAP...08..015A}. Likewise, neutrino primaries interact directly in the ice and produce radio emission solely through the Askaryan mechanism. As a result, the in-ice radio emission from cosmic-ray showers can closely resemble that induced by neutrinos~\cite{Kockere1.2022PhRvD.106d3023D}. We also note that ray-tracing solutions for the in-ice emission toward deep in-ice antennas should yield two distinct paths: a direct path and an indirect path where the ray reflects at the ice–air interface. Yet, for mildly inclined down-going showers such as those considered in this article, geometrical configurations leading to an observable indirect emission are unlikely.

In both air and ice emissions, the radio signal is emitted predominantly within a cone around the shower axis, with a maximum intensity at the so-called Cherenkov angle, as illustrated by the purple conical region in Fig.~\ref{fig:CRsketch}. This Cherenkov compression arises from the relativistic propagation of the shower particles, which causes the radiation emitted from different points along the shower axis to arrive in phase at the Cherenkov angle~\cite{Cherenkov2012APh....37....5W,KrijnCherenkov2013APh....45...23D}. As a result, the radio signal on the Cherenkov cone is highly coherent, extending to higher frequencies and exhibiting a rapid spatial variability.

\section{Building a cosmic ray library}~\label{sec:FAERIE}
\subsection{FAERIE simulation framework}
While commonly used Monte-Carlo codes such as ZHAireS~\cite{ZHS2012APh....35..325A} and CoREAS~\cite{COREAS2013AIPC.1535..128H} can simulate the in-air component of cosmic ray radio emission and propagate it down to the Earth's surface, these tools cannot currently propagate the emission below the ice nor generate the in-ice component, as the rapid variation of the ice density with depth bends the emission and requires to implement ray tracing. To simulate radio emission in such a complex medium, recent developments aim, for example, to couple the CORSIKA8 framework with the Eisvogel simulation tool~\cite{Corsika8:2025LT,Windischhofer:2023Jn}.  On the other hand, FAERIE~\cite{FAERIE2024PhRvD.110b3010D} is a recently developed numerical tool, that combines the CORSIKA~\cite{Corsika1998cmcc.book.....H} and GEANT4~\cite{GEANT42016NIMPA.835..186A} Monte-Carlo codes, coupled to the endpoint formalism~\cite{endpointPhysRevE.84.056602} and a ray tracing algorithm, to simulate both the in-air and the in-ice radio emission from cosmic ray particle cascades, as seen by deep in-ice observers~\cite{FAERIE_ARENA2024arXiv240902185C, Chiche:2025yy}. In this work, we  use the FAERIE framework to investigate in-ice radio signatures of cosmic ray particle showers and perform the first characterization of their emission in a dense inhomogeneous medium.

\begin{figure*}[tb]
\includegraphics[width=0.49\linewidth]{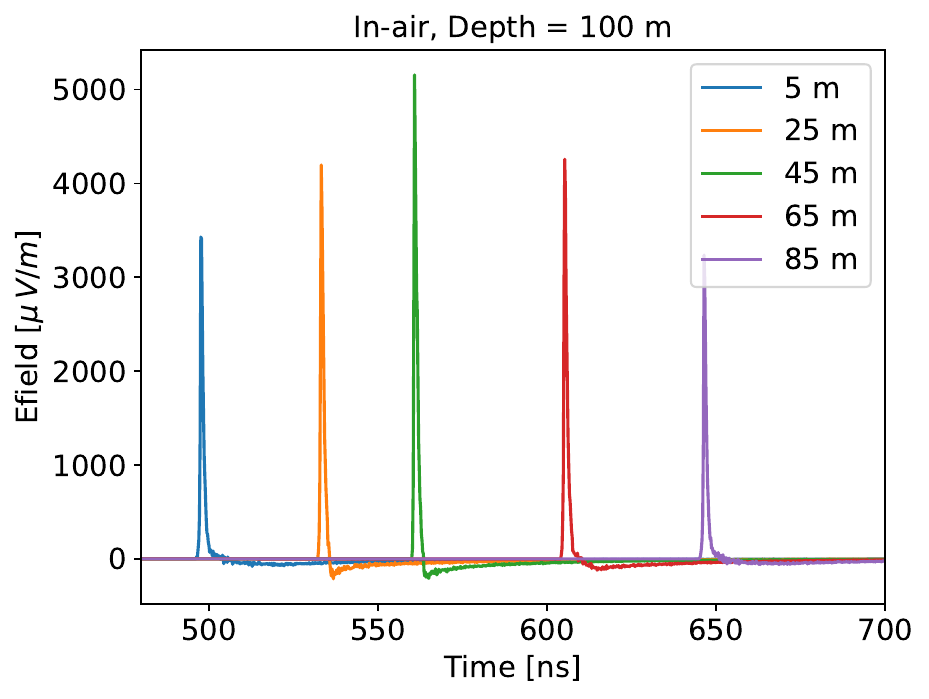}\hfill
\includegraphics[width=0.49\linewidth]{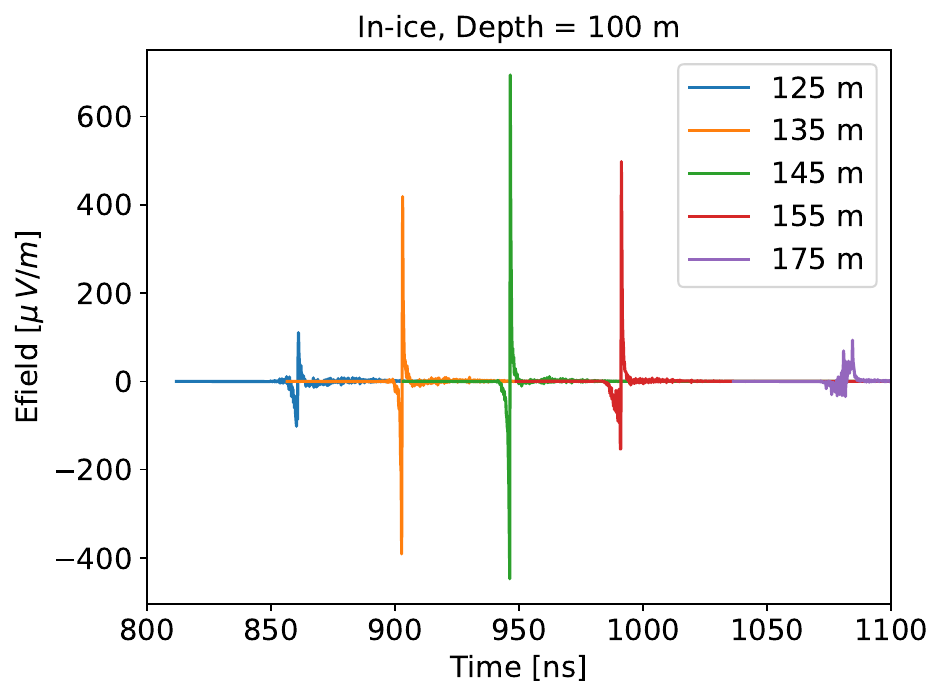}
\caption{East-West component of the electric field simulated with FAERIE for the in-air ({\it left}) and in-ice ({\it right}) component of cosmic ray radio emission for a proton-induced shower with primary energy $E=10^{17.5}\, \rm eV$ and zenith angle $\theta=34^{\circ}$. Antennas are located at a depth of 100 m and placed at various radial distances from the shower core.}\label{fig:traces}
\end{figure*}

Running a given FAERIE simulation requires to give as inputs the primary particle nature (mass composition), energy and arrival direction (zenith and azimuth angle). The user must also specify a list of antenna positions that can be above or below the Earth's surface (i.e., the air/ice boundary) at which the emission is evaluated. The outputs are the three components electric field time traces at each antenna for both the in-air component ($E_x^{\rm air}(t)$, $E_y^{\rm air}(t)$, $E_z^{\rm air}(t)$) and the in-ice component ($E_x^{\rm ice}(t)$, $E_y^{\rm ice}(t)$, $E_z^{\rm ice}(t)$) stored independently. We note that the in-ice traces incorporate both direct and indirect solutions. The outputs are given in the CoREAS reference frame, with $x$: North-South, $y$: West-East, and $z$: Up-Down. In Fig.~\ref{fig:traces} we illustrate outputs of FAERIE simulations for the in-air  (left-hand panel) and in-ice (right-hand panel) components by showing the East-West channel of the electric field time traces for antennas located at a depth of 100 meters below the ice and at various radial distances from the shower core. The typical cosmic ray radio emission corresponds to a sharp pulse with a width of a few nanoseconds and an amplitude of a few to thousands of $\mu {\rm V/m}$. Pulses from both components exhibit a bipolar structure, as the electric field is proportional to the time derivative of the vector potential, which increases during the shower development and decreases as it vanishes.
The figure also shows that the signal varies with the antenna positions depending on their radial distance to the  shower core and in particular their relative position with respect to the Cherenkov cone which is found for this specific shower at a radial distance $r_{\rm cer}^{\rm air}\sim 45\, \rm m$ and $r_{\rm cer}^{\rm ice}\sim 145\, \rm m$ for the in-air and in-ice components respectively.


\subsection{Simulation strategy}~\label{sec:Setup}

Using FAERIE, we simulated the radio emission of cosmic ray showers with different primary energy and arrival directions, to build a library of events and perform the first characterization of their emission as seen by deep in-ice observers. FAERIE is a computationally expensive tool, with a single simulation that can run for $\sim10^{4}-10^{5}$ CPU hours at EeV energies, which corresponds to approximately 3 weeks of calculation when run in parallel on the local T2B HTCondor cluster~\cite{t2b}. As a consequence, to build our library of events we ran only one shower per set of primary particle energy and zenith angle. While this does not allow us to fully characterize effects related to shower-to-shower fluctuations, this still allows us to capture the main global characteristics of cosmic ray radio emission and their dependencies on  the shower parameters. Furthermore, we mitigated our results by running 10 identical showers with a primary energy $E=10^{16.5}\, \rm eV$ and zenith angle $\theta=0^{\circ}$ to estimate the uncertainties in our results related to shower-to-shower fluctuations. This however motivates the need for the development of faster simulations, while not losing too much on accuracy.

\begin{table}[b]
\centering
\begin{tabular}{|c|c|}
\hline
\textbf{Energy [eV]} \rule{0pt}{2.5ex} & $10^{16.5}$; $10^{17}$; $10^{17.5}$ \\
\hline
\textbf{Zenith [°]} & 0; 20; 28; 34; 39; 43; 47; 50 \\
\hline
\textbf{Depth [m]} & 0; 60; 100 \\
\hline
\end{tabular}
\caption{Primary energy, zenith bins and depths of the FAERIE simulation set.}
\label{table:setup}
\end{table}

\begin{figure*}[tb]
\includegraphics[width=0.51\linewidth]{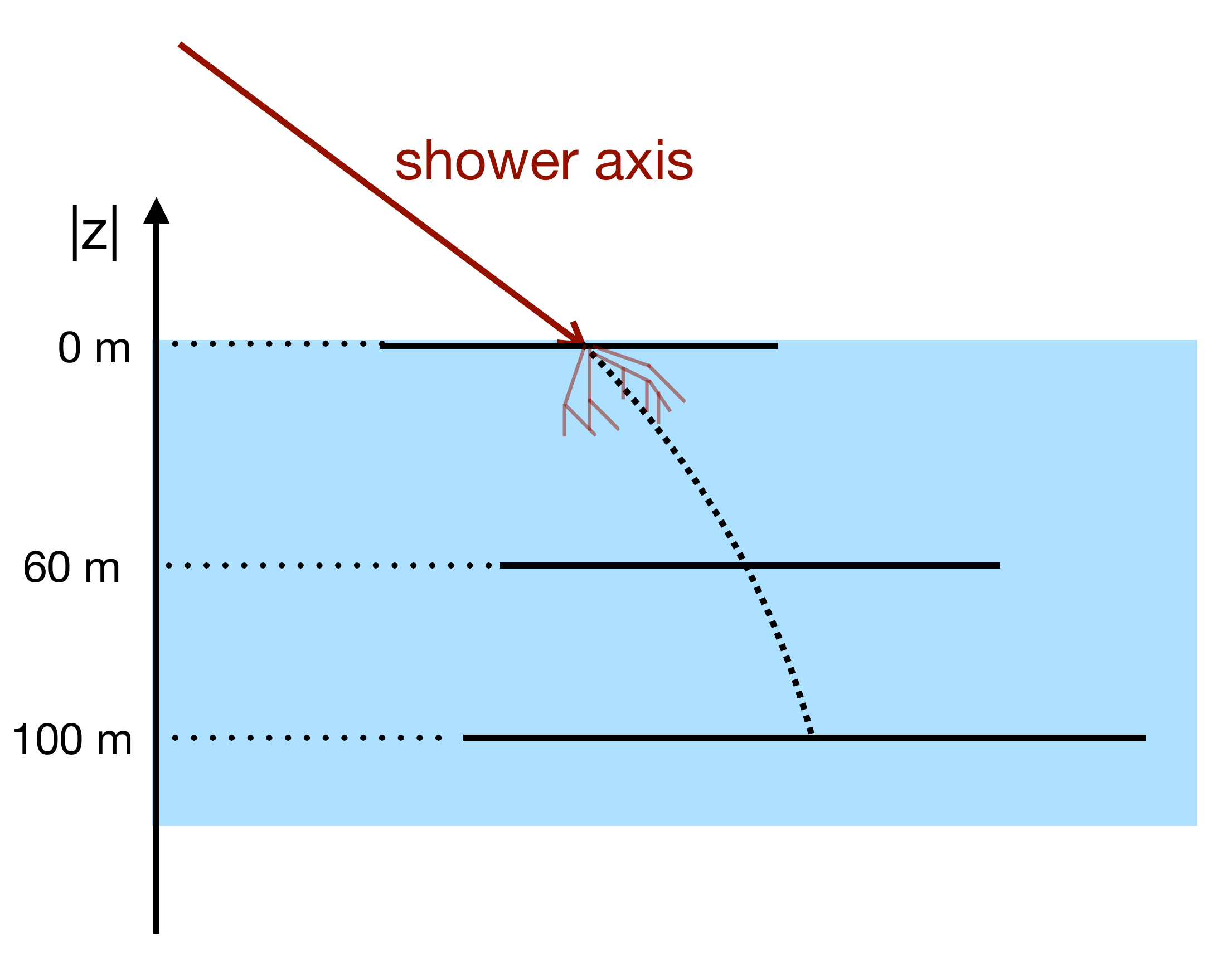}\hfill
\includegraphics[width=0.46\linewidth]{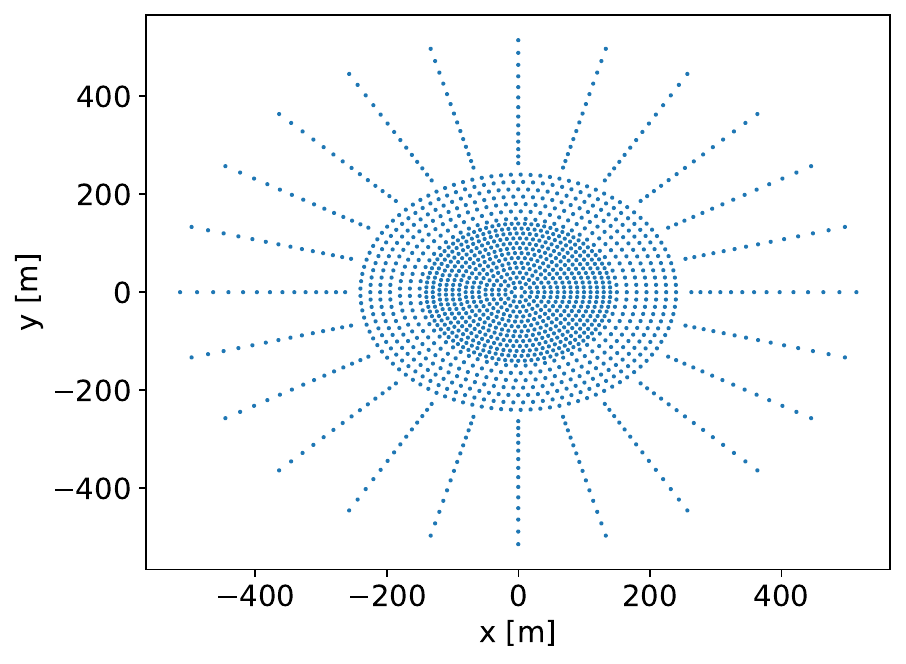}
\caption{({\it Left}) Sketch of antenna layers at various depths  below the ice surface. The red arrow represents the shower axis, the black dotted curved line shows the in-ice continuation of the shower axis using Snell-Descartes law. Each antenna layer is centered on its intersection with the dotted line. The layer width increases with depth. ({\it Right}) Typical antenna layout for a given layer. The central region is a dense core with 10-m spacing out to 150 m, followed by 15-m spacing up to 250 m. Beyond this, the array extends into 24 radial arms separated by $15^{\circ}$, with antennas placed along each arm at logarithmically increasing radial distances. Each given layer has 1568 antennas, the full grid corresponds to 4704 simulated antennas.}\label{fig:AntennaLayout} 
\end{figure*}

\subsection{Shower input parameters}~\label{sec:inputs}
For our library, we simulated, using QGSJETII-04 as hadronic model, 24 proton showers with parameters given in Table~\ref{table:setup}. We used the default GEANT configuration of the FAERIE installation which includes the following physics lists~\cite{FAERIE2024PhRvD.110b3010D, deKockere_IceShelf}: G4EmStandardPhysics, G4DecayPhysics and G4RadioactiveDecayPhysics.
We note that hadronic interactions are not simulated for the development of the in-ice cascade in the current setup. However, secondary hadronic interactions induced by electrons and positrons from the cascade are expected to have only a small impact on the shower development and the resulting effect on the radio emission was shown to be of a few percents or less~\cite{Kockere1.2022PhRvD.106d3023D}.

We considered three primary particle energies generated in logarithmic scale in the range $[10^{16.5}-10^{17.5}]\, \rm eV$. This range was chosen as showers with energies equal or below $E=10^{16}\, \rm eV$ will hardly trigger the detector and as the statistics of showers with energies equal or above $E=10^{18}\, \rm eV$ is too low to contribute significantly to cosmic ray detections. Though at ultra-high energy the cosmic ray composition is expected to be mainly heavy~\cite{MassCompAuger2017JCAP...04..038A}, we simulated only proton primaries as their larger shower-to-shower fluctuations should provide us with results that encompass the typical signals expected for heavier primaries such as iron. This choice was also motivated by the fact that a few test iron simulations run with FAERIE does not show significant differences with analogous proton simulations. At first order, cosmic ray radio emission should scale  with the cosine of the shower zenith angle, hence  zenith angles were generated uniformly in $\cos{\theta}$  between $\theta  = 0^{\circ}$ and $\theta  = 50^{\circ}$. We did not simulate more inclined showers since (1) no in-ice component of cosmic ray emission is expected for very inclined showers as they fully develop in-air (see Section~\ref{fig:RadEvszen}), and (2) FAERIE currently relies on a flat-Earth description, an assumption that breaks when considering  very inclined showers which can propagate in the atmosphere for hundreds of kilometers. The simulation library was designed for a typical Greenland in-ice detector such as RNO-G, hence the magnetic field was set to the one of Summit Station, Greenland, where RNO-G is located, with amplitude $B_x = 7.7 \,\mu {\rm T}$, $B_z =-54.1 \, \mu {\rm T}$. This magnetic field being near-vertical, we considered only one azimuth angle for our library of events ($\varphi=0^{\circ}$, shower propagating towards the magnetic North) as we assume that the radio emission does not vary significantly with the shower azimuth angle (apart from a rotation and a shift of the radio footprints).  All our simulations were run with a relative thinning of $10^{-4}$, a value chosen to achieve a compromise between the precision of the simulations and the computation time. Although this value is relatively high, comparisons with a few showers simulated with a thinning of $10^{-5}$ show very similar features. Nevertheless, such a thinning may still lead to a slight overestimation of the uncertainties related to shower-to-shower fluctuations. All the results shown in this study are filtered in the $50-1000\, \rm MHz$ band to match the realistic range of observed signals.

\subsection{Ice density model}~\label{sec:IceModel}
To run  a FAERIE simulation we must also specify the ice refractive index profile. We used an  exponential ice model based on measurements of the ice properties in Greenland, following~\cite{IceModel2018PhRvD..98d3010D}:

\begin{equation}
    n(z) = A - B\exp({-C|z|}) \ ,
\end{equation}
where $n$ is the ice refractive index, $|z|$ is the depth of the antennas and $A$, $B$, $C$ are parameters of the model.  Greenland ice has a discontinuity around a depth of $|z|=14.9\, \rm m$ due to a change in the ice compactification. Hence the refractive index is best fitted using a double exponential model, i.e., two sets of parameters depending on the depth. Following~\cite{IceModel2018PhRvD..98d3010D} we used, $A = 1.775$, $B =0.5019$, $C =0.03247\, \rm m^{-1}$ for $|z| < 14.9{\, \rm m}$ and $A = 1.775$, $B =0.448023$, $C =0.02469\, \rm m^{-1}$ for $|z| > 14.9{\, \rm m}$.

\begin{figure*}[tb]
\includegraphics[width=0.50\linewidth]{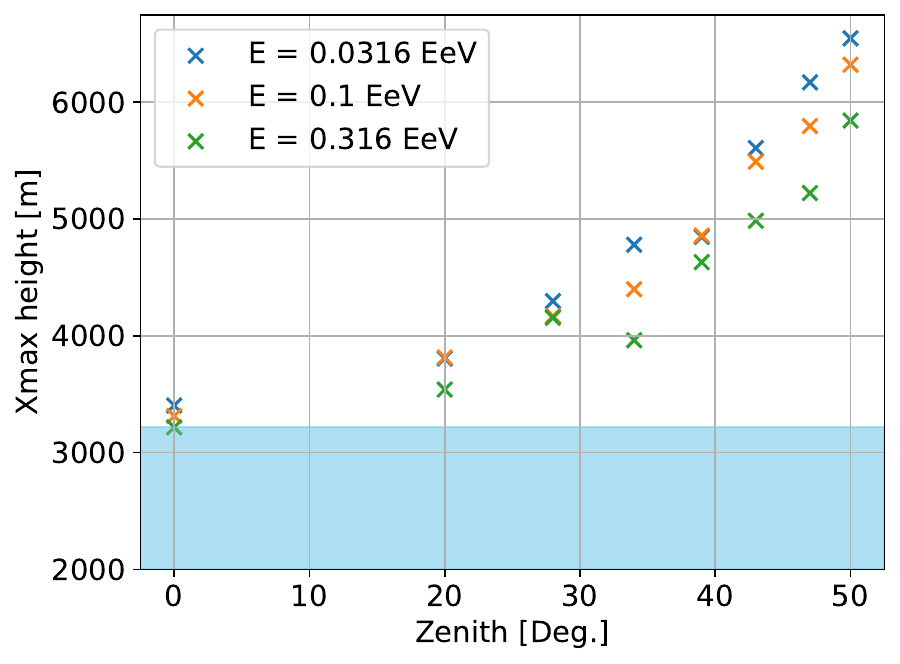}\hfill
\includegraphics[width=0.467\linewidth]{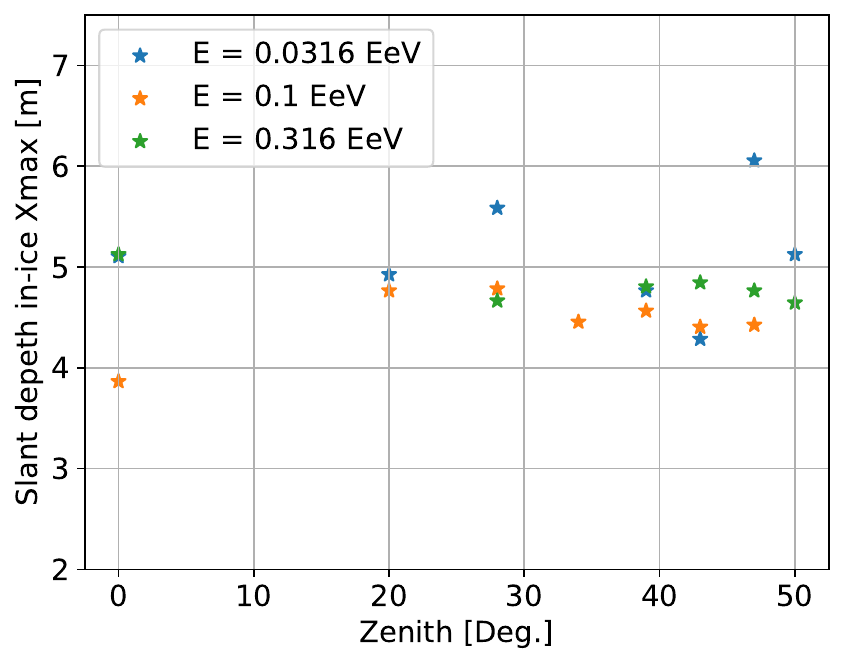}
\caption{({\it Left}) Height above sea level of the in-air shower maximum $X_{\rm max}$ from FAERIE simulations, as a function of the shower zenith angle, for different primary energies (colors).  The blue shaded area indicates the ice surface. ({\it Right}) Slant depth (along the shower axis), of the in-ice shower maximum, as a function of the shower zenith angle, for the same simulation set.}\label{fig:XmaxCharac}
\end{figure*}

\subsection{Antenna grid}~\label{sec:Grid}
We considered a 3D grid of 4704 antennas corresponding to three layers of 1568 positions each, located  at a depth of $0\, \rm m$ (surface), 60 meters and 100 meters, as illustrated in the left-hand panel of Fig.~\ref{fig:AntennaLayout}. These depths were chosen to match those of the RNO-G detector, but are consistent with other detectors like ARA or the future IceCube-Gen2 radio detector. For the same reason, the air-ice boundary is set at an altitude of 3216 meters, corresponding to the one of Summit Station in Greenland. A typical antenna layer is shown in the right-hand panel of Fig.~\ref{fig:AntennaLayout}. We used a polar grid of antennas with a dense circular core between 0 and 250 meters from the shower impact point, and 24 outer arms, evenly spaced by $15^{\circ}$ each, for larger radial distances. The spacing of the inner core is fixed for every simulation, while the spacing of the outer arms is logarithmic and increases with the depth of the antenna layer and the shower zenith angle, to account for the fact that the footprint size increases with the distance to the emission source. This specific grid was designed to capture both the specificities of the radio emission from the in-ice component, which requires a dense grid independently of the shower parameters, and those of the in-air component, which requires a sparse zenith-dependent spacing (see discussion Section~\ref{sec:LDF}). Furthermore, as also shown in Fig.~\ref{fig:AntennaLayout}, we shifted the antenna layers below the surface laterally so that our grid is always centered on the radio emission footprints. The offset was calculated by propagating a ray along the shower axis below the surface using Snell-Descartes's law. Deep layers are hence horizontally displaced from the shower core ($x= 0$, $y=0$), because of this offset.

\subsection{Distribution of shower maximum}~\label{sec:ShowerMax}
\begin{figure}[ht]  
    \centering
    \includegraphics[width=\linewidth]{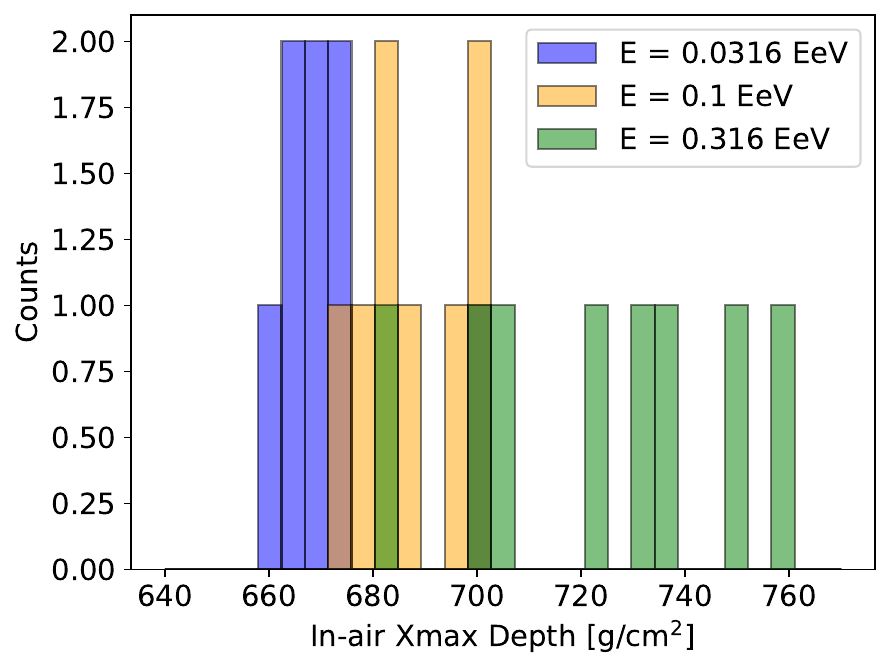}
    \caption{Distributions of the in-air shower maximum grammage $X_{\rm max}$, for different primary energies.}
    \label{fig:XmaxDepth}
\end{figure}

 We characterized our simulation library by evaluating the shower maximum $X_{\rm max}$ of the in-air and the in-ice cascades, defined as the depth at which the longitudinal profile of the total particle number integrated over the full transverse extent of the shower core, reaches its maximum. The left-hand panel in Fig.~\ref{fig:XmaxCharac}, shows the height above sea level of the in-air $X_{\rm max}$ as a function of the shower zenith angle. As expected, inclined showers have their $X_{\rm max}$ at higher altitude than vertical ones since the inclined showers propagate longer in the upper atmosphere. We further note that at fixed zenith angle, more energetic showers usually have a deeper $X_{\rm max}$ height, since the $X_{\rm max}$ grammage scales logarithmically with the primary energy~\cite{Heitler2005APh....22..387M, XmaxEngel:2011zzb}. Eventually, we observe that for vertical showers, with a zenith angle $\theta=0^{\circ}$, $X_{\rm max}$ can even be below the ice surface. This implies that in such cases the in-air cascade is not yet fully developed when it reaches the ground, leading to clipping effects~\cite{GlaserRadE2016JCAP...09..024G}. Hence, in all the further results that we will show the case of vertical showers should be treated with care and analyzed separately because of this effect.  In the right-hand panel of Fig.~\ref{fig:XmaxCharac}, we show the slant depth of the in-ice $X_{\rm max}$, i.e., the distance of the in-ice $X_{\rm max}$ measured along the shower axis starting from the impact location. We note that in practice, only particles within  a radius of 1 meter in the plane perpendicular to the shower axis, at the shower impact location, are propagated into the ice. Here, we observe that all simulated showers have their in-ice $X_{\rm max}$ at  around the same distance from the shower core, independently of the primary particle energy or zenith angle. This is a striking feature as it points towards universality of the in-ice cores induced by cosmic ray particle cascades. In Fig.~\ref{fig:XmaxDepth}, we show the $X_{\rm max}$ grammage of our simulations. Although the $X_{\rm max}$ should not depend on the shower zenith angle, we can observe variations of the grammage even at fixed primary energy which are related to shower-to-shower fluctuations. Still, we get a mean $X_{\rm max}$ grammage of $\sim 670\, \rm g/cm^{2}$ at $10^{16.5}\, \rm eV$ and of $\sim 740\, \rm g/cm^{2}$ at $10^{17.5}\, \rm eV$, which are typically expected values for proton simulations with QGSJETII-04.

\section{Spatial amplitude distribution}~\label{sec:Amplitude}
\subsection{Radio emission footprints}~\label{sec:footprints}

\begin{figure*}[tp]
    \centering

    \begin{minipage}[c]{\textwidth}
    \subcaption*{(a) In-air emission, $\theta = 0^\circ$}
        \centering
        \begin{minipage}[c]{0.32\textwidth}
            \centering
            \includegraphics[width=\linewidth]{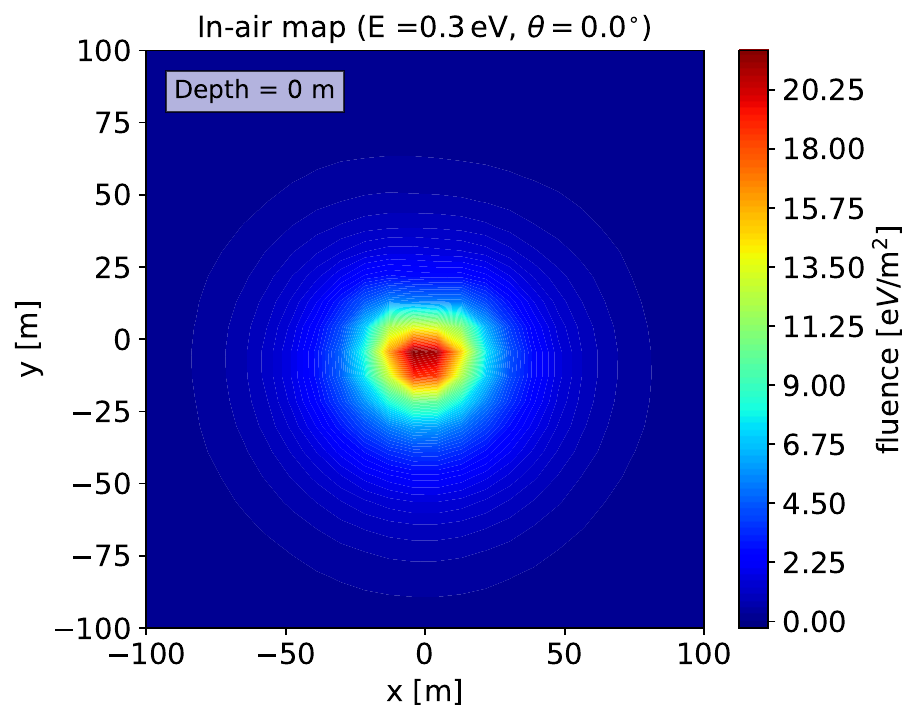}
        \end{minipage}\hfill
        \begin{minipage}[c]{0.32\textwidth}
            \centering
            \includegraphics[width=\linewidth]{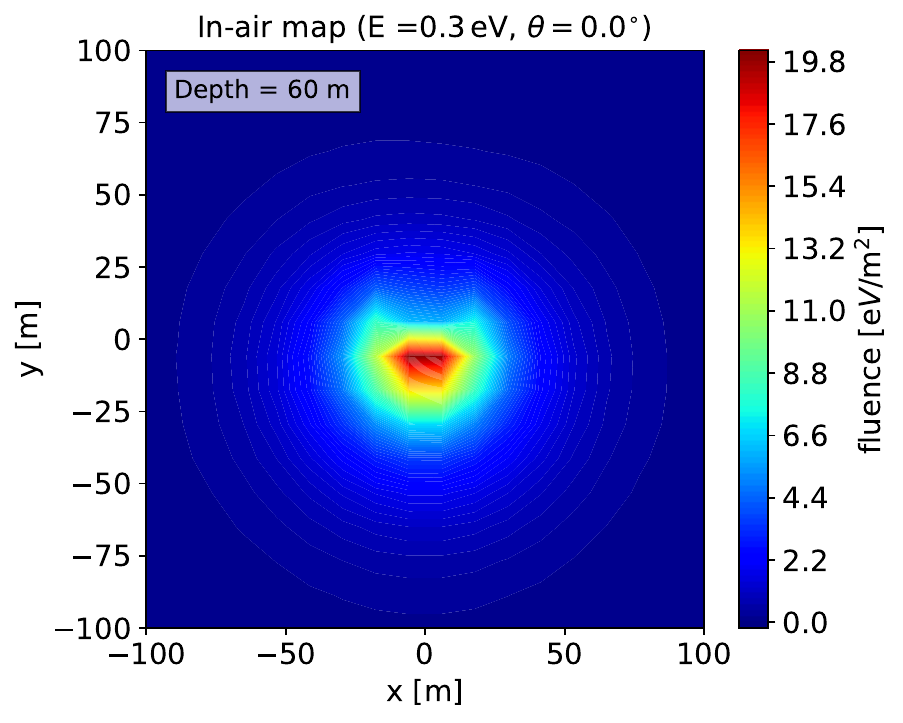}
        \end{minipage}\hfill
        \begin{minipage}[c]{0.32\textwidth}
            \centering
            \includegraphics[width=\linewidth]{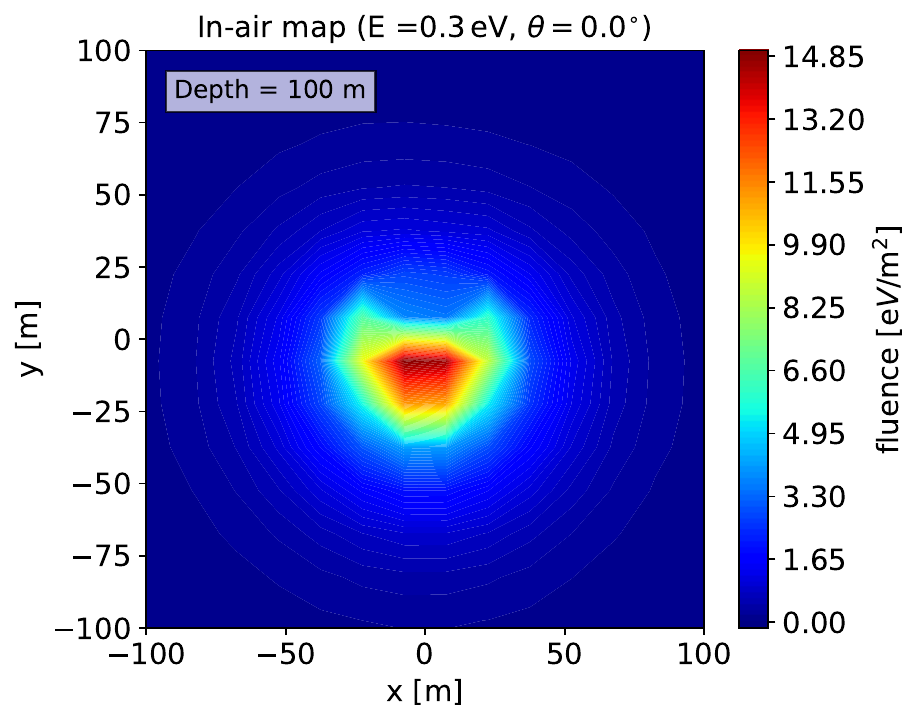}
        \end{minipage}

    \end{minipage}

    \vspace{0.4cm}

    \begin{minipage}[c]{\textwidth}
    \subcaption*{(b) In-ice emission, $\theta = 0^\circ$}
        \centering
        \begin{minipage}[c]{0.32\textwidth}
            \centering
            \includegraphics[width=\linewidth]{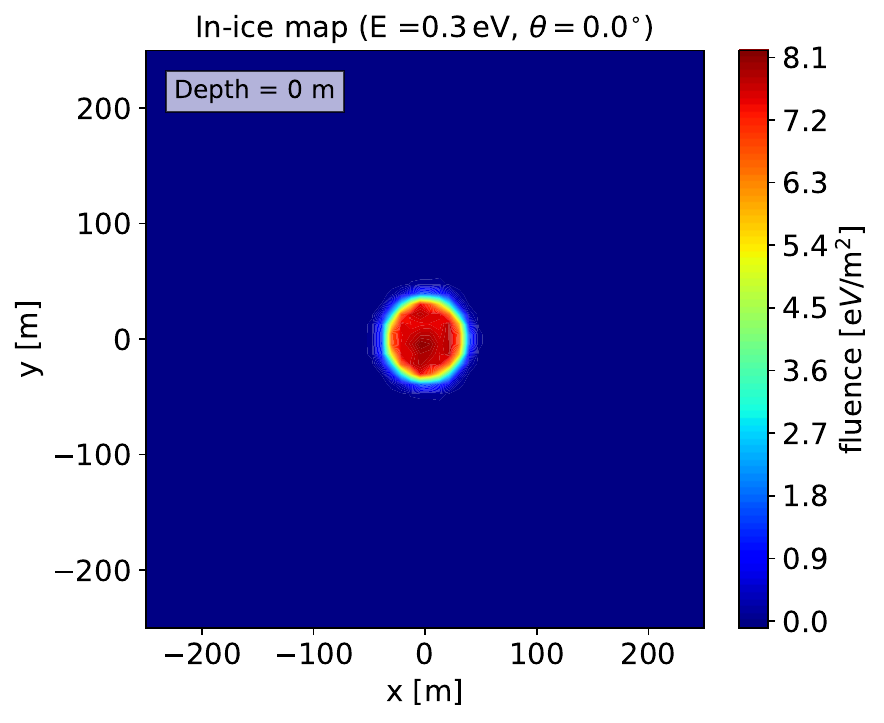}
        \end{minipage}\hfill
        \begin{minipage}[c]{0.32\textwidth}
            \centering
            \includegraphics[width=\linewidth]{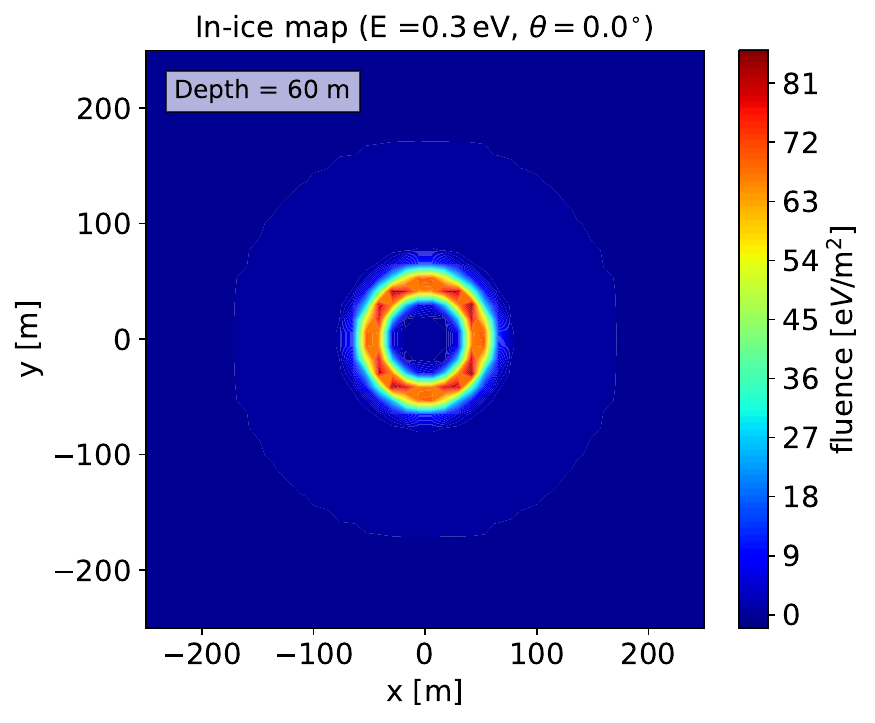}
        \end{minipage}\hfill
        \begin{minipage}[c]{0.32\textwidth}
            \centering
            \includegraphics[width=\linewidth]{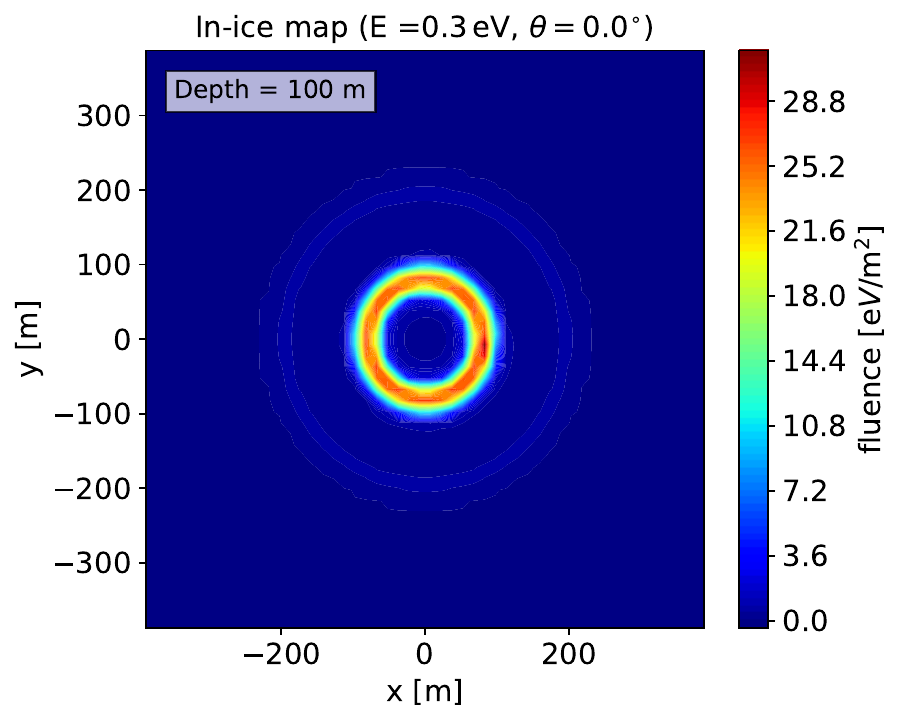}
        \end{minipage}

    \end{minipage}

    \vspace{0.6cm}

    \begin{minipage}[c]{\textwidth}
    \subcaption*{(c) In-air emission, $\theta = 50^\circ$}
        \centering
        \begin{minipage}[c]{0.32\textwidth}
            \centering
            \includegraphics[width=\linewidth]{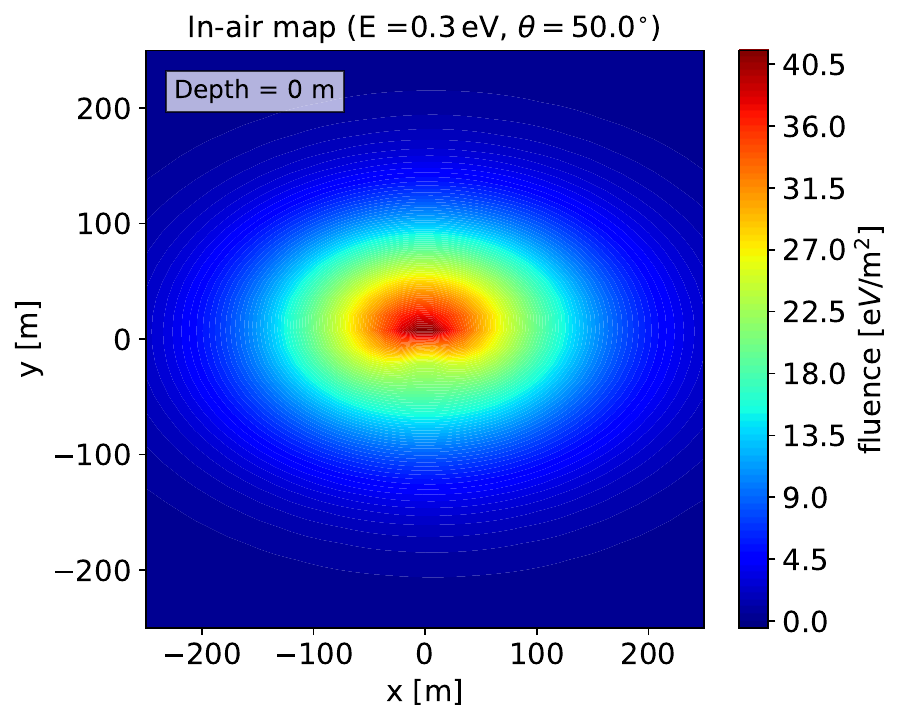}
        \end{minipage}\hfill
        \begin{minipage}[c]{0.32\textwidth}
            \centering
            \includegraphics[width=\linewidth]{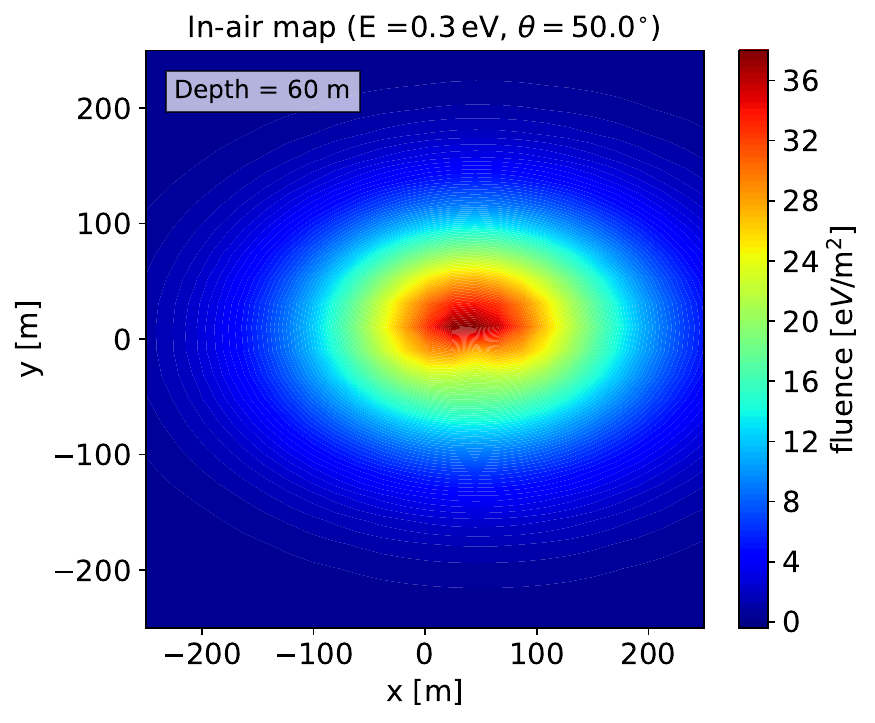}
        \end{minipage}\hfill
        \begin{minipage}[c]{0.32\textwidth}
            \centering
            \includegraphics[width=\linewidth]{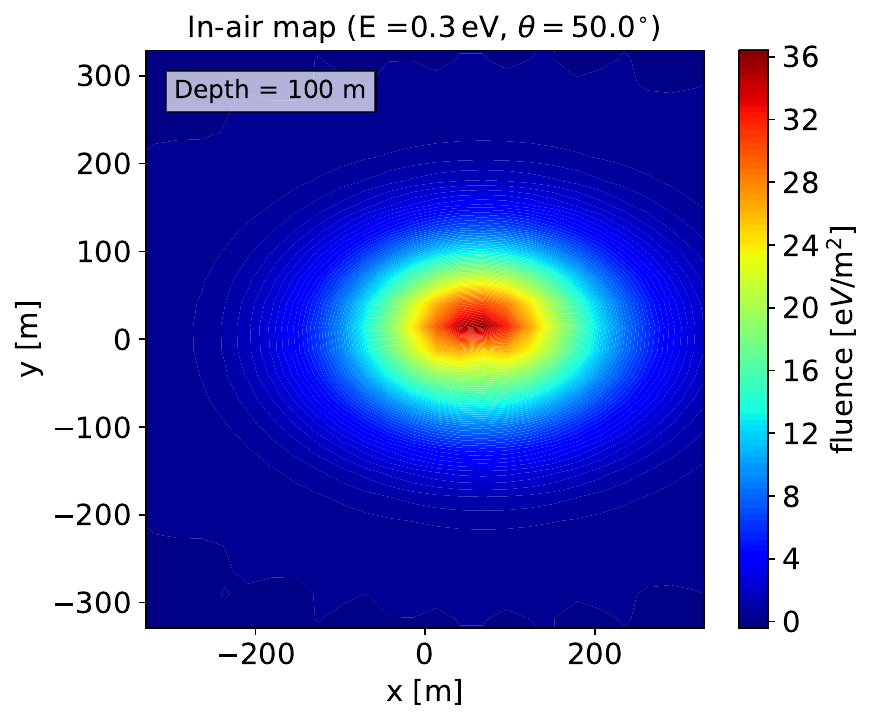}
        \end{minipage}

    \end{minipage}

    \vspace{0.4cm}

    \begin{minipage}[c]{\textwidth}
     \subcaption*{(d) In-ice emission, $\theta = 50^\circ$}
        \centering
        \begin{minipage}[c]{0.32\textwidth}
            \centering
            \includegraphics[width=\linewidth]{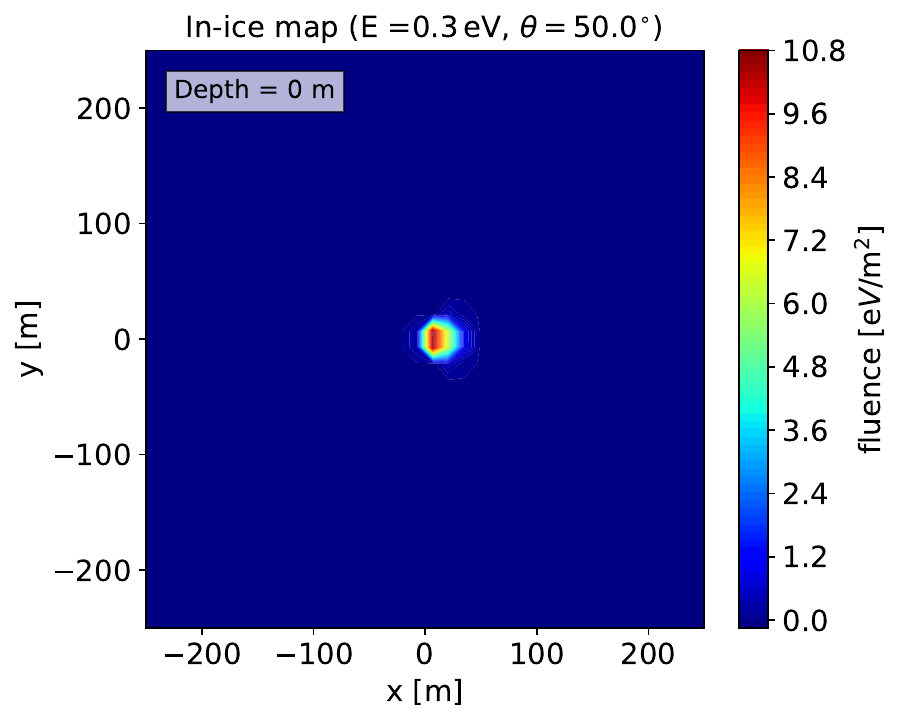}
        \end{minipage}\hfill
        \begin{minipage}[c]{0.32\textwidth}
            \centering
            \includegraphics[width=\linewidth]{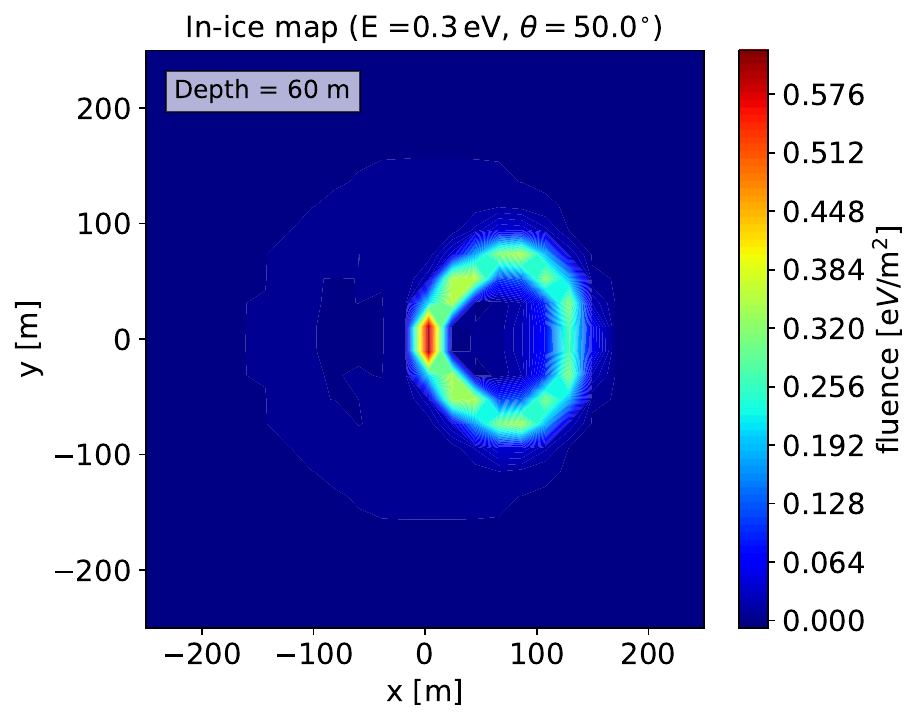}
        \end{minipage}\hfill
        \begin{minipage}[c]{0.32\textwidth}
            \centering
            \includegraphics[width=\linewidth]{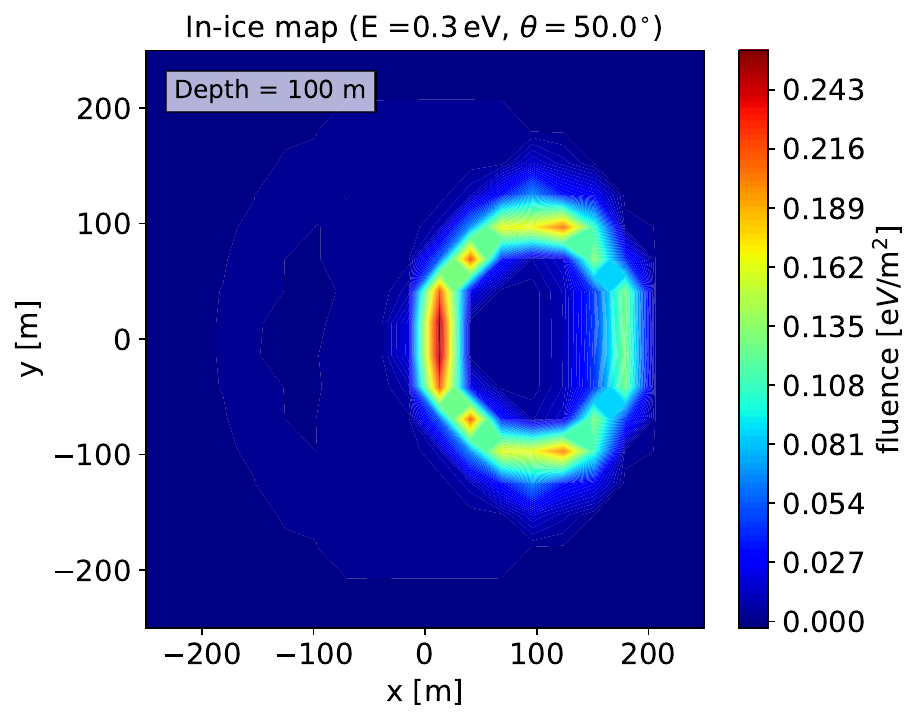}
        \end{minipage}

    \end{minipage}

    \caption{
        Interpolated electric-field maps of in-air and in-ice emission for proton showers at 
        $10^{17.5}\,\mathrm{eV}$, at zenith angles $0^\circ$ and $50^\circ$, and for antenna depths of 
        0, 60 and 100 m. Each row corresponds to a different shower geometry, and each mini-caption 
        summarizes the conditions.
    }\label{fig:footprints}
\end{figure*}

Using FAERIE, we evaluated the radio footprints of our simulations as a function of the antenna depths and the shower parameters. For this, we derived the energy fluence of the radio emission at each antenna, which measures the energy per unit area and is given by integrating over time the squared electric field following~\cite{GlaserRadE2016JCAP...09..024G}

\begin{equation}
    f_{x,y,z}(x,y) = \epsilon_0 c\int_{t_{\rm min}}^{t_{\rm max}}E_{x,y,z}^{2}(t,x,y) \,{\rm d}t \ ,
    \label{eq:fluence}
\end{equation}
where $\epsilon_0$ is the vacuum permittivity and $c$ is the speed of light in vacuum and where the integration is performed over the full time window, with bounds $t_{\rm min}$ and $t_{\rm max}$. The total energy fluence at each antenna is then derived from the sum of its three components $f_{\rm tot} = f_x + f_y + f_z$. In Fig.~\ref{fig:footprints}, we show interpolated footprints of the total energy fluence from the in-air and in-ice components for a vertical proton shower with zenith angle $\theta=0^{\circ}$ [lines (a) and (b) respectively] and for a shower with $\theta=50^{\circ}$ [lines (c) and (d)] and primary energy $E=10^{17.5}\, \rm eV$. The radio signal is interpolated using a two-dimensional radial basis function method with a linear kernel, ensuring a smooth reconstruction over the antenna grid even for a grid with non-uniform spacing. The results are shown for the three different depths,  $0,\,60,$ and 100 meters from left to right. The results are shown here for only one primary energy, but we note that lower primary energies overall yield similar results with lower energy fluence at the antennas.

Looking first at the in-air emission from the vertical shower [line (a)], it can be seen that the radio footprint follows a bean-shaped pattern with a stronger emission on the negative $y$-axis and a weaker emission on the positive $y$-axis. This feature is well-known and expected from the interferences between the geomagnetic and charge-excess mechanisms. Indeed, as detailed in Section~\ref{sec:Emission}, the geomagnetic emission is linearly polarized toward the $-\mathbf{v} \times \mathbf{B}$ direction while the charge-excess has a radial polarization. This implies that on one side of the $\mathbf{v} \times \mathbf{B}$ axis positive interferences between both mechanisms are expected, while on the other side destructive interferences are expected, leading to this bean-shaped pattern. We also notice that no clear Cherenkov features are visible for the in-air emission which is expected  when the Cherenkov angle is small and the observer is close to $X_{\rm max}$. This is typically the case for showers in Greenland due to the high altitude above sea level which yields an in-air $X_{\rm max}$ very close to the ground, as shown in Fig.~\ref{fig:XmaxCharac}. In line (b) we show the in-ice emission of the same shower. The maps at a depth of 60 and 100 meters show a clear rotationally symmetric high amplitude Cherenkov ring structure. We also note that while the in-ice emission should be focused in a cone around the shower axis, the map of surface antennas (depth = 0 m) shows a non-negligible emission very close to the shower core ($x=0$, $y=0$). This feature was not reported before but may be expected as the in-ice cascade develops very close to the ice surface, which implies that the surface lies within the emission region. While the amplitude of this component appears too low to be detectable, the reliability of its modeling in this near-surface regime is uncertain. We therefore do not consider this contribution in the following and only investigate the in-ice radio emission for deep antennas.

When the showers are inclined, both the in-air and the in-ice emissions are modified, as shown by lines (c) and (d). A first notable change for both emissions is that for inclined showers, the radio footprints of deep antenna layers are no longer centered on the shower core ($x=0$, $y=0$) but shifted towards the shower propagation direction, corresponding to the positive $x$-axis here. For the in-air emission [line (c)], we also note that the footprints are elongated towards the shower propagation direction, which is only a geometrical effect linked to the position of the in-air shower maximum with respect to the ice surface.  For the in-ice emission [line (d)], we notice that the radio footprint is no longer rotationally symmetric but deformed. This deformation increases with depth and is linked to the bending of the radio emission as it propagates in the ice, due to the varying refractive index of the medium. We also note that the radio emission amplitude is larger on the side of the Cherenkov cone closest to x=0 than on the opposite side. This asymmetry originates from the fact that radiation from the two sides of the Cherenkov cone travels over different path lengths to the antennas, leading to different observed amplitudes.

Comparing the amplitude of the different footprints, we observe that for the vertical shower the maximal amplitude of the in-ice emission is higher than that of the in-air emission, while for the inclined shower the in-ice emission amplitude gets very weak and much lower than the one of the in-air footprint, a feature that will be further discussed in Section~\ref{fig:RadEvszen}. Looking at the evolution of the footprint amplitude with depth, we also notice that when going to deeper layers, the amplitude of the in-ice emission decreases rapidly, by a factor 2-3. On the other hand, the amplitude of the in-air emission decreases but much less significantly with almost no variation between a depth of $|z|=60$ meters and $|z|=100$ meters for the inclined shower. This effect is purely geometrical. The in-ice $X_{\rm max}$ is closer to the observer than the in-air $X_{\rm max}$, so the relative distance to the emission source varies more rapidly with depth for the in-ice component. We also mention that the amplitude variation with depth for the in-air component is expected to decrease with increasing shower zenith angle, as the shower maximum gets further away from the observer (Fig.~\ref{fig:XmaxCharac}). This geometric effect has an important implication: the in-ice signal exhibits a rapid and characteristic variation of amplitude with depth, whereas radio sources located far above the surface would produce a much smoother depth dependence. Such a distinct depth profile could therefore be exploited to reject distant anthropogenic backgrounds, such as airplanes~\cite{airplanesRNOG2025JInst..20P1015A}, and to identify in-ice cosmic-ray or neutrino signals.

 Finally, we mention that for the in-ice emission of the vertical shower observed at a depth of 100 meters [right-hand panel in line (b)] the footprint shows a low-amplitude double ring structure. This feature appears consistently in our simulations when the energy of the in-ice cascade is sufficiently large. Its physical origin is not yet understood. One possibility is that the outer ring is produced by an additional emission component from the in-ice particle cascade, emitted at angles larger than the Cherenkov angle. This interpretation is motivated by the fact that the corresponding polarization pattern seems similar to the radial polarization expected for Askaryan emission. Ray-tracing calculations indicate that, if this emission originates near the in-ice $X_{\rm max}$, signals reaching antennas located at large horizontal distances ($r\sim 200\, \rm m$) would require near-horizontal launch angles, with $i_{\rm launch}\sim86^{\circ}$. We also note that this feature may be related to an interference pattern observed in the frequency domain, as discussed in Section.~\ref{sec:frequency}. Other mechanisms may also contribute. For instance, transition radiation emitted at the air–ice interface could in principle produce an additional emission component, although it is currently unclear whether it would lead to the observed footprint and polarization pattern. The origin of this second ring therefore remains an open question and will be investigated in future studies.  If confirmed, this could be a distinctive signature of in-ice cosmic ray radio emission that could be used for their identification. Yet, the amplitude of this component being very low, it may only be detectable for sufficiently energetic particle cascades.

\begin{figure*}[t]
    \centering

    \begin{minipage}{0.45\textwidth}
        \centering
        \includegraphics[width=\linewidth]{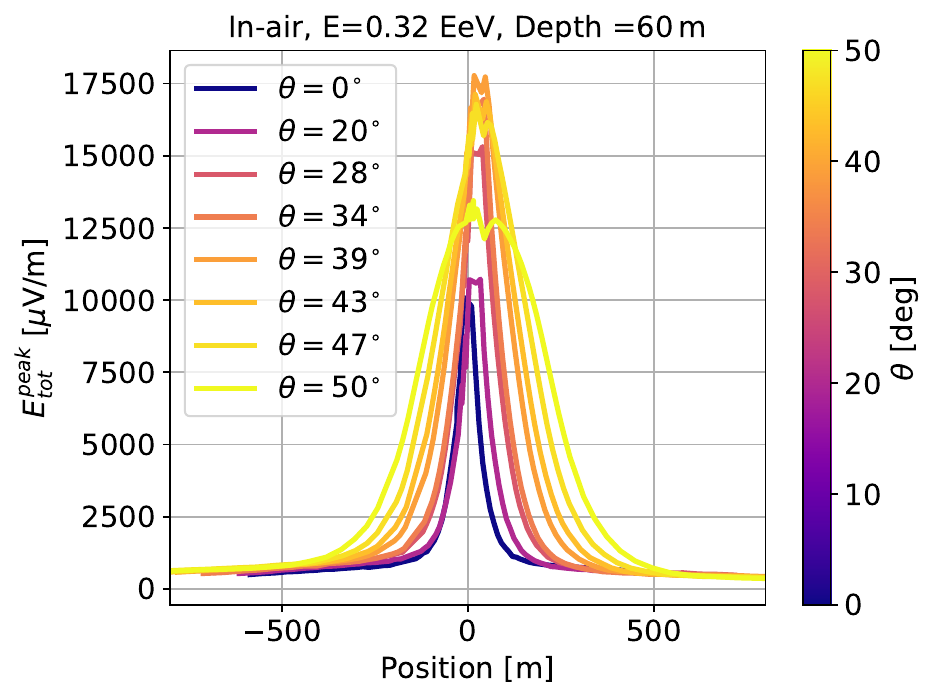}
    \end{minipage}
    \hfill
    \begin{minipage}{0.45\textwidth}
        \centering
        \includegraphics[width=\linewidth]{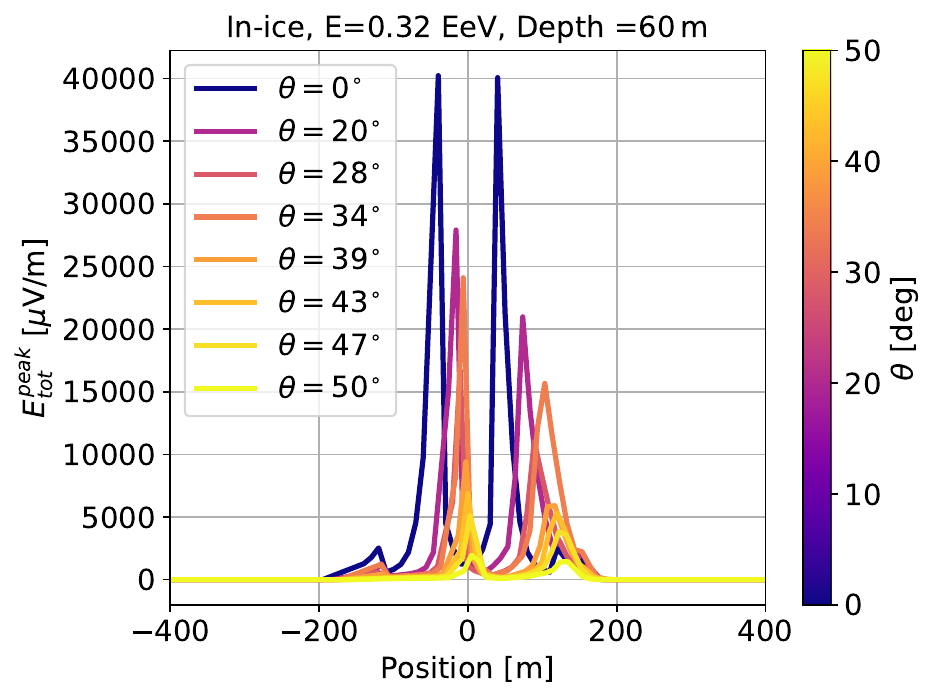}
    \end{minipage}

    \vspace{0.4cm}

    \begin{minipage}{0.48\textwidth}
        \centering
        \includegraphics[width=\linewidth]{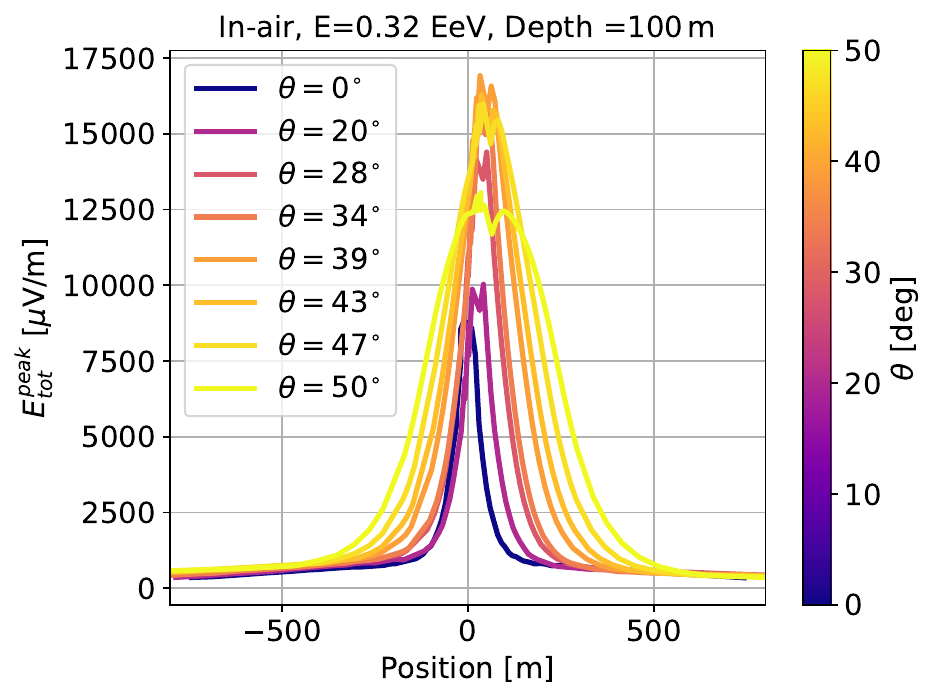}
    \end{minipage}
    \hfill
    \begin{minipage}{0.48\textwidth}
        \centering
        \includegraphics[width=\linewidth]{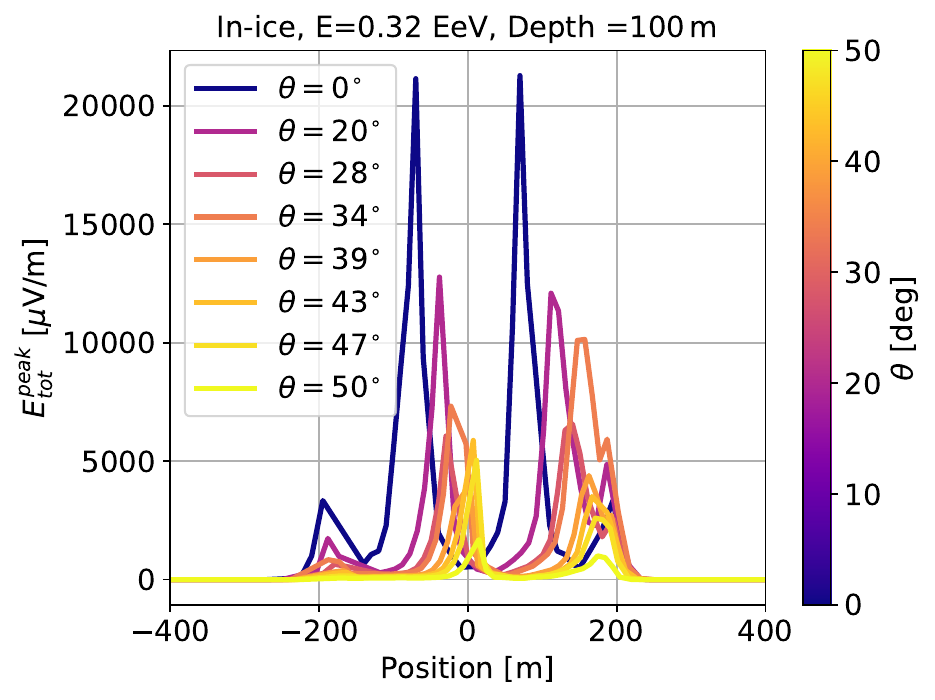}
    \end{minipage}

    \caption{Lateral distribution functions of the peak amplitude of the electric-field norm along the South-North axis, observed at a depth of 60 meters ({\it top}) and 100 meters ({\it bottom}), for the in-air ({\it left}) and in-ice ({\it right}) emissions. Simulated  showers with primary energy $E=10^{17.5}\, \rm eV$ were used. Each colored line corresponds to a different zenith angle.}
    \label{fig:LDF}
\end{figure*}

\subsection{Lateral distribution functions}~\label{sec:LDF}

To characterize the evolution of the radio footprint spatial dependency with the shower parameters, we also evaluated the lateral distribution function (LDF) of the emission, which corresponds to the electric field peak amplitude along a given baseline of antennas. In Fig.~\ref{fig:LDF} we show the LDF of the in-air (left-hand panels) and the in-ice (right-hand panels) emissions, along the $x$ (North-South) axis, at a depth of $60\,  \rm m$ (top panels) and $100 \, \rm m$ (bottom panels), for cosmic ray showers with a primary energy $E= 10^{17.5}\, \rm eV$ and various zenith angles.

As mentioned in the previous section, it is striking on these plots that the variation of the signal amplitude with depth is more important for the in-ice component than for the in-air one. These plots also highlight the minor Cherenkov feature for the in-air emission and the distinctive two peaks for the in-ice component. We also notice that when increasing the zenith angle, the in-ice emission amplitude decreases and becomes asymmetric. Interestingly the small peak related to the second ring observed in Fig.~\ref{fig:footprints} seems to be always at the same position around $\pm 190 \, \rm m$, even when the shower gets more inclined. Finally, we point that independently of the shower zenith angle the in-ice radio emission footprint  does not extend over more than $\pm 200\, \rm m$, which justifies the choice of the dense core in our antenna grid in Fig.~\ref{fig:AntennaLayout}.

\begin{figure*}[tb]
\includegraphics[width=0.49\linewidth]{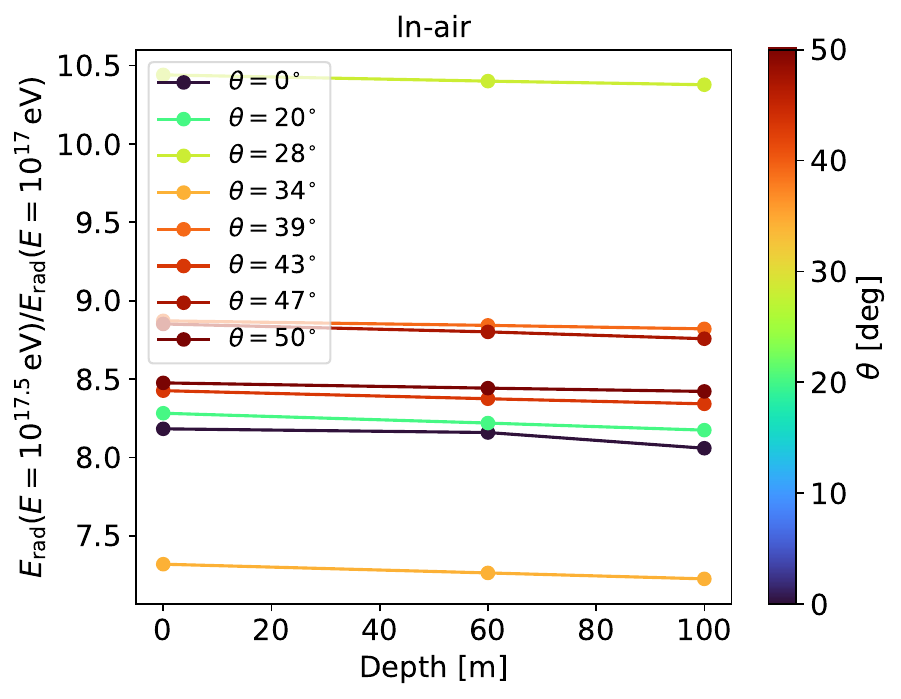}\hfill
\includegraphics[width=0.49\linewidth]{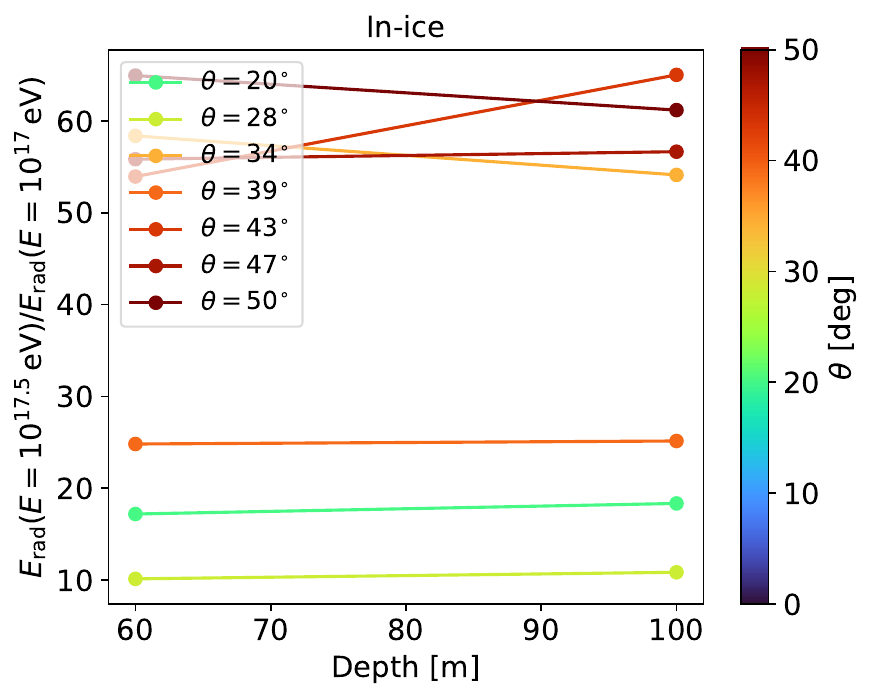}
\caption{ Radiation energy ratio for the in-air ({\it left}) and in-ice ({\it right}) emissions, between showers with primary energy $E=10^{\rm 17.5}\,\rm eV$ and $E= 10^{\rm 17}\,\rm eV$, for different zenith angles (colors), as a function of the antenna depth. For the right-hand plot, the vertical case ($\theta=0^{\circ}$) is not shown as it corresponds to a specific case where the shower at $E=10^{17.5}\, \rm eV$ has its in-air $X_{\rm max}$ below the ice.}\label{fig:RadEcheck}
\end{figure*}

For the in-air emission, two effects are in competition when the shower gets inclined. First, the shower maximum $X_{\rm max}$, i.e., the point-like source of the emission becomes further away from the antenna grid which implies that the LDF extends over larger distances but also that the radio emission amplitude decreases accordingly by energy conservation. On the other hand, inclined showers have their in-air $X_{\rm max}$ higher in the atmosphere (Fig.~\ref{fig:XmaxCharac}), leading to a stronger transverse current due to the lower density in which the shower develops and a higher radiated energy from the geomagnetic mechanism~\cite{GlaserRadE2016JCAP...09..024G}(see also Section~\ref{sec:footprints}). This explains why intermediate zenith angles tend to have the LDF with the highest maximal amplitude, as seen here for $\theta=39^{\circ}$.

\section{Radiation energy}~\label{sec:RadE}

In order to constrain global characteristics of cosmic ray radio signals we evaluated the radiation energy of the emission. The radiation energy is obtained by integrating the energy fluence over a given antenna layer, following~\cite{GlaserRadE2016JCAP...09..024G}:
\begin{equation}
    E_{\rm rad} = \int_{x_{\rm min}}^{x_{\rm max}}\int_{y_{\rm min}}^{y_{\rm max}}f(x,y)\,{\rm d}x\, {\rm d}y \ .
    \label{eq:RadE}
\end{equation}
By definition, the radiation energy is a calorimetric observable which quantifies the amount of energy which is radiated in terms of radio emission. Its dependency with the shower parameters can therefore be directly linked to the shower physics. In the following we evaluate and compare the radiation energy of the in-air and in-ice components of cosmic ray radio emission.

\subsection{Consistency checks}~\label{sec:RadEchecks}

The radiation energy is sensitive to the grid of antennas used to evaluate the radio signal and can be biased if the antenna grid is too sparse, which would deteriorate the interpolation of the radio signal, or if the grid is not large enough, in which case part of the radio footprint could be missing. Since we simulated a 3D grid of antennas, a simple consistency check consists in evaluating the radiation energy using different layers of antennas independently. As the radiation energy is conserved through propagation of the emission, for a single shower the radiation energy should be roughly the same independently of the layer used for its computation.  In Fig.~\ref{fig:RadEcheck}, we show the square root of the ratio of the radiation energy between showers with primary energy $E=10^{17.5}\, \rm eV$  and $E=10^{17}\, \rm eV$, for different zenith angles. The left-hand panel corresponds to the in-air component while the right-hand panel is for the in-ice component. We note that for the in-ice emission we only included antenna layers at 60 and 100 meters since these are the most relevant ones to characterize the emission. We can observe that for a fixed zenith angle, the radiation energy is constant across all depths for both the in-air and the in-ice emission. Only a minor deviation is observed for the in-ice component at a zenith angle of $\theta=43^{\circ}$, but the variation still remains small. This implies that our simulation library can safely be used to characterize the radiation energy of the radio emission.

\subsection{Energy scaling}~\label{sec:RadEscaling}

From Fig.~\ref{fig:RadEcheck}, we can also study the energy scaling of the radio emission. 
An increase in the primary energy leads, to first order, to a proportional increase in the number of particles at the in-air shower maximum~\cite{Heitler2005APh....22..387M}, such that $N_{\rm part} \propto E_{\rm primary}$. For coherent radio emission, the radiation energy then scales with the square of the number of emitters, $E_{\rm rad}^{\rm air} \propto N^{2}_{\rm part}$, implying $\sqrt{E_{\rm rad}^{\rm air}} \propto E_{\rm primary}$.  Since the radiation energy is defined as the time integral of the squared electric field, the electric-field amplitude must therefore scale linearly with the primary energy, $\mathcal{E} \propto E_{\rm primary}$. 

\begin{figure}[ht]  
    \centering
    \includegraphics[width=\linewidth]{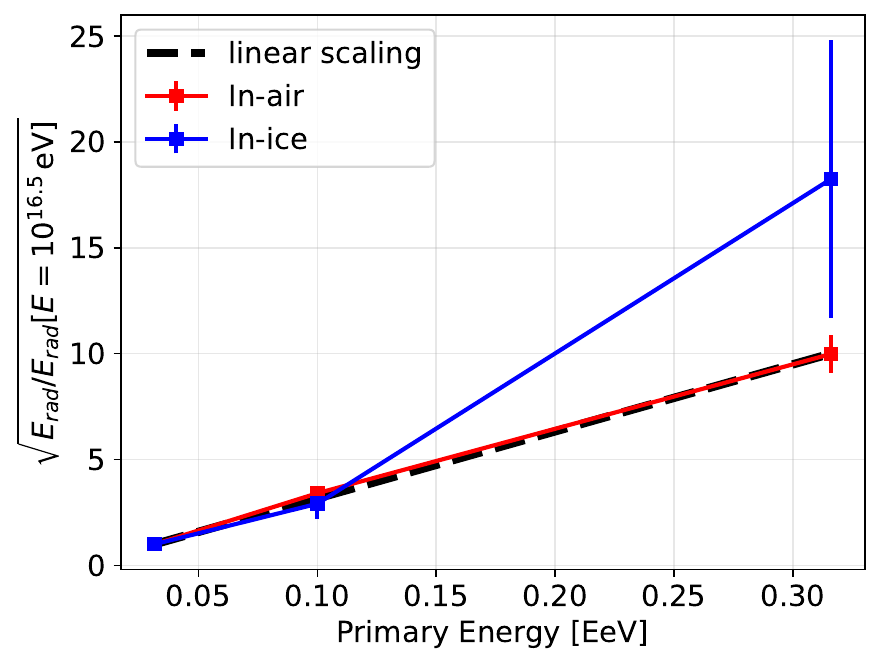}
    \caption{Square root of the in-air (red) and in-ice radiation energies, as a function of the primary particle energy and normalized by the value at $10^{16.5}\, \rm eV$. Each data point corresponds to average over the 8 zenith angles of our library, while the error bars show the standard deviations of the distributions.}
    \label{fig:EnergyScaling}
\end{figure}

\begin{figure*}[tb]
\includegraphics[width=0.49\linewidth]{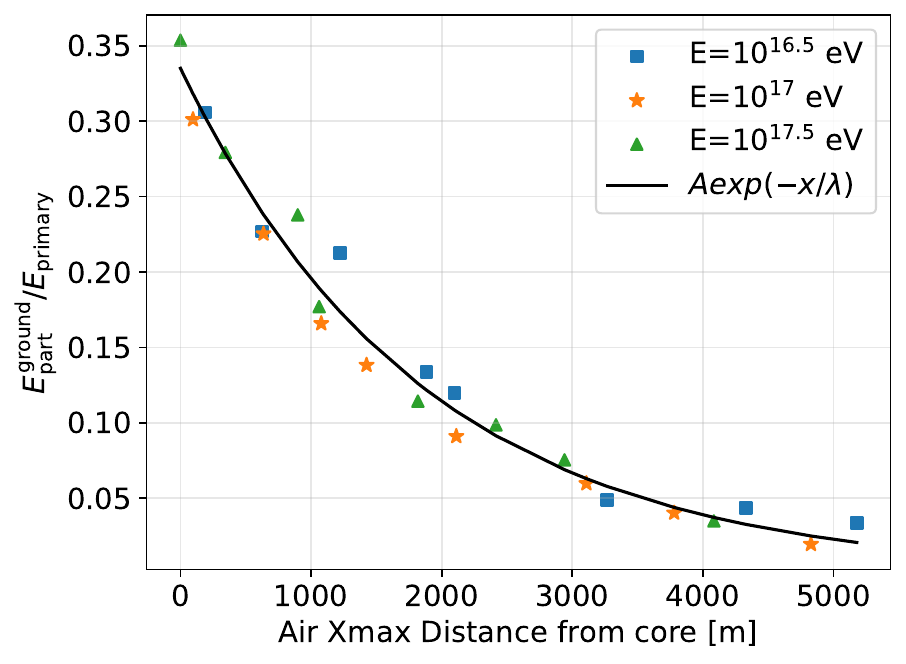}\hfill
\includegraphics[width=0.49\linewidth]{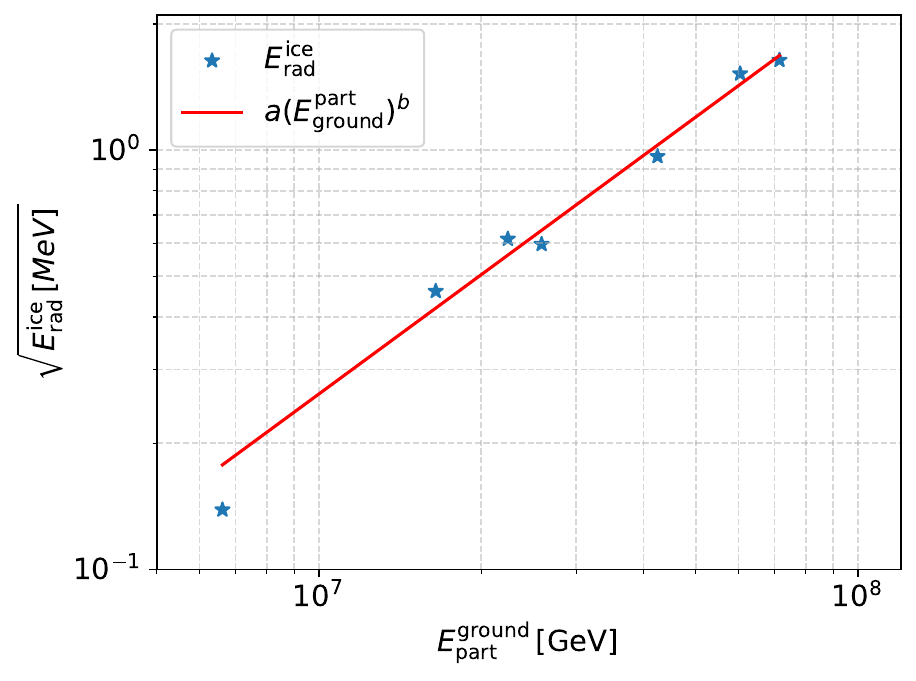}
\caption{({\it Left}) Total ground particle energy as a function of the distance from the in-air shower maximum  $X_{\rm max}$. The color and marker style codes for the various primary energies. The results are fitted with an exponential decay law (black line). ({\it Right}) Square root of the in-ice radiation energy as a function of the ground particle energy for simulated showers with primary energy $E=10^{17.5}\, \rm eV$. Each data point corresponds to a shower with different zenith angle. The results are fitted with a linear model following $\sqrt{E^{\rm ice}_{\rm rad}}=a(E_{\rm ground}^{\rm part})^b$, which yields $a=2.4\times10^{-8}\pm 1\times 10^{-9}$, a slope $b=0.945 \pm 0.06$ and a coefficient of determination $R^2 = 0.99$.}\label{fig:GroundPartScaling}
\end{figure*}

These scaling relations can be directly confirmed in Fig.~\ref{fig:RadEcheck}. The ratio of the two primary energies considered is $10^{17.5}/10^{17}=\sqrt{10}$, which implies a factor of $\sim10$ in radiation energy. This is indeed approximately what is observed for the in-air component in Fig.~\ref{fig:RadEcheck}. The fact that the values are on average slightly below 10 can be attributed to the fact that in practice the radiation energy rather scales with the electromagnetic component of the shower than with the total number of particles. On the other hand, the in-ice emission shows ratios well above 10 for some showers, which indicates that for this component, $\sqrt{E_{\rm ice}^{\rm rad}}$ grows faster than linearly with the primary energy. This trend is further evidenced in Fig.~\ref{fig:EnergyScaling}, which shows the square root of the in-air and in-ice radiation energies as a function of the primary particle energy and the comparison to a linear scaling. Each data point on the figure corresponds to an average over the 8 zenith angles of our library while the error bars represent the standard deviations of the distributions. Before averaging, the radiation energy at each zenith was normalized to its value at $10^{16.5}\, \rm eV$, so as to remove any dependence on the shower zenith angle. We observe that the in-air emission indeed follows a linear scaling while a super-linear scaling is observed for the in-ice component, particularly at the highest energies. The large error bars at high energy also indicate large variations of the in-ice energy scaling depending on the shower zenith angle.

To better constrain the energy scaling of the in-ice emission, we can decompose the problem into two steps. (1) We first evaluate the total energy of the particles reaching the ice surface, $E_{\rm part}^{\rm ground}$, and its dependence on the shower parameters. (2) We then relate $E_{\rm part}^{\rm ground}$ to the in-ice radiation energy $E^{\rm ground}_{\rm part}$. Indeed, following the same reasoning as for the in-air emission, the in-ice radiation energy should scale with the number of particles at the ice surface,  $E^{\rm ice}_{\rm rad} \propto (E_{\rm part}^{\rm ground})^{2}$, rather than directly with the primary particle energy. The  particle energy at the ice surface is expected to depend both on the shower primary particle energy and on the geometrical distance between the in-air $X_{\rm max}$ and the ice surface, since shower particles will lose their energy as they propagate in the atmosphere. We remove the first dependency by normalizing $E^{\rm ground}_{\rm part}$ to the primary energy, while we show the remaining dependency with the distance to $X_{\rm max}$, obtained from FAERIE simulations, in left-hand panel of Fig.~\ref{fig:GroundPartScaling}. We can observe that the normalized particle energy at the ice surface is in good agreement with an exponential decay   following
\begin{equation}
    \frac{E^{\rm ground}_{\rm part}}{E_{\rm primary}} = A\exp\left(-\frac{x}{\lambda}\right) \ ,
\end{equation}
with $A$ a proportionality factor and $\lambda \sim 1800 \, \rm m$ the characteristic attenuation length of the particle cascades. The distance $D_{\rm xmax}$ of the shower core to the in-air $X_{\rm max}$, itself depends only on the Earth's atmosphere and on the primary particle characteristics. Indeed, the primary particle mass composition and energy will determine its $X_{\rm max}$~\cite{XmaxEngel:2011zzb} grammage, light energetic primaries usually penetrate deeper in the atmosphere than heavier/less energetic ones. Then given a $X_{\rm max}$ grammage, an atmospheric density model and the shower zenith angle, one can derive the $X_{\rm max}$ position and its distance to the Earth's surface.  

\begin{figure*}[tb]
\includegraphics[width=0.49\linewidth]{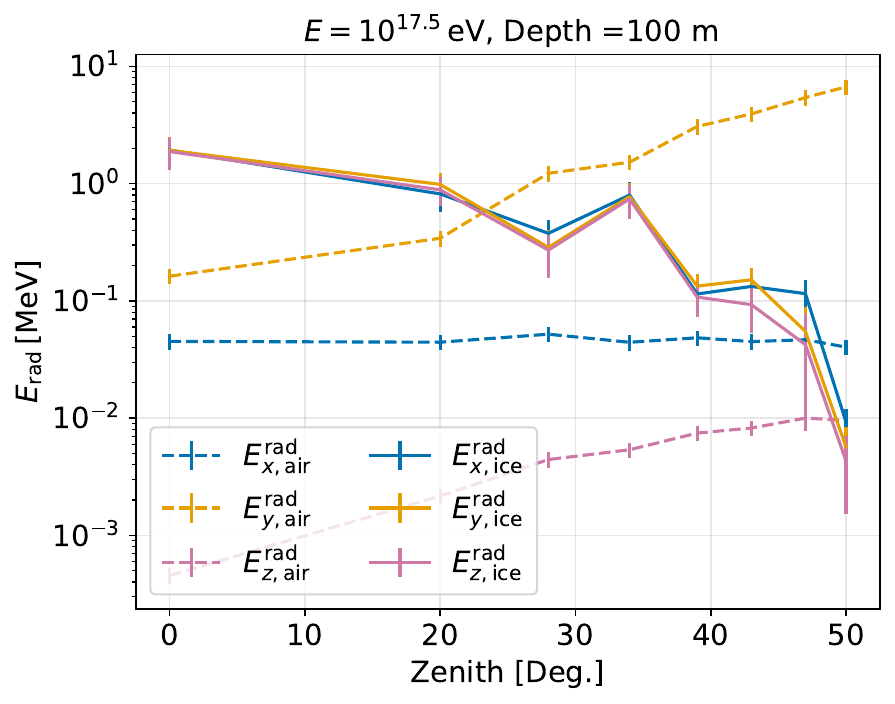}\hfill
\includegraphics[width=0.49\linewidth]{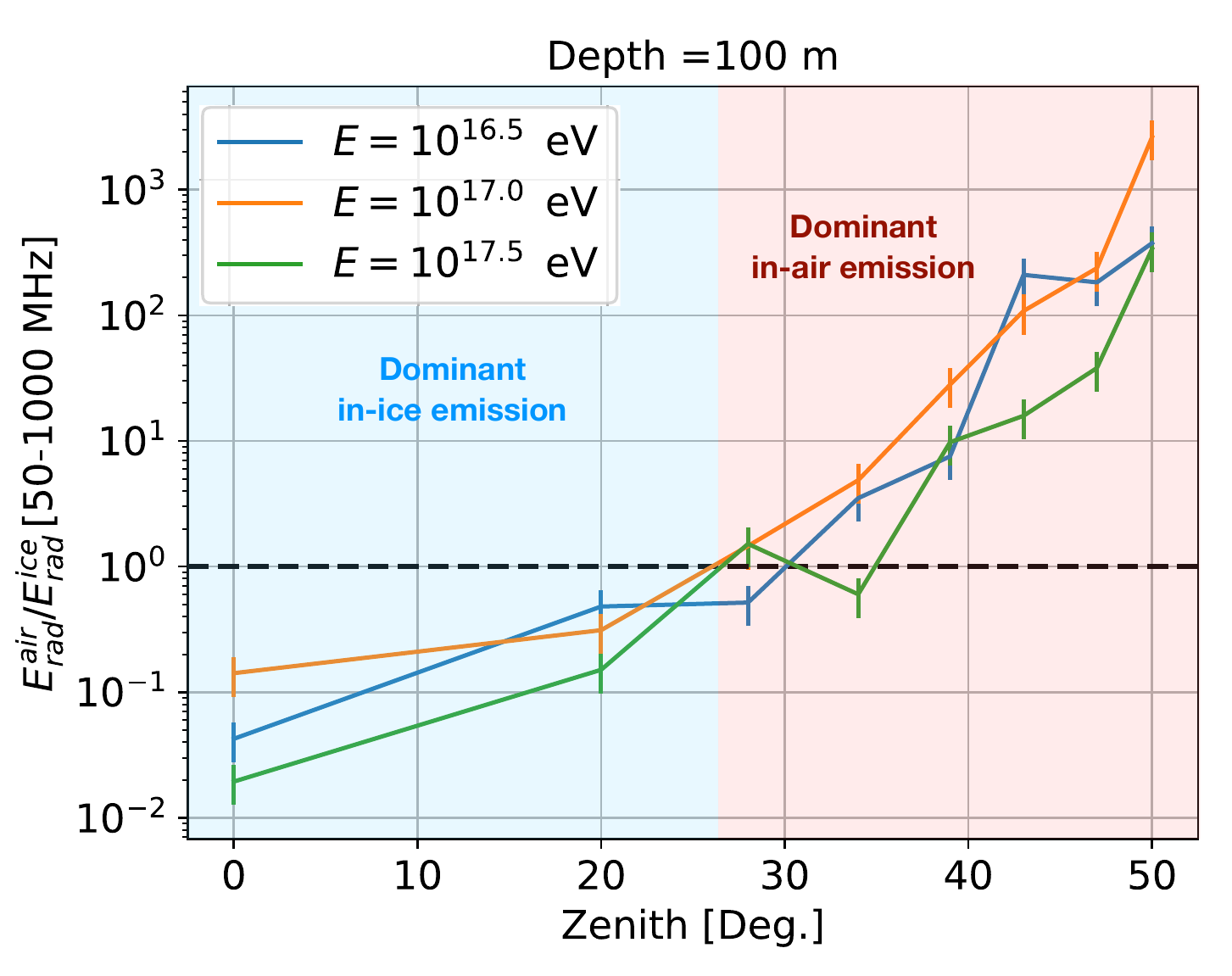}
\caption{({\it Left}) In-air (dashed line) and in-ice (solid lines) radiation energy associated with each channel of the electric field, as a function of the shower zenith angle. The results correspond to simulations with primary energy $E=10^{\rm 17.5}\, eV$ and the antenna layer at depth of $100\, \rm m$ was used to evaluate the radiation energy. ({\it Right}) Ratio between the in-air and in-ice radiation energies, as a function of the shower zenith angle, for three primary particle energies (colored lines). The black dashed line indicates the limit where the ratio is equal to 1. On both plots the error bars represent the estimated uncertainties related to shower-to-shower fluctuations.}\label{fig:RadEvszen}
\end{figure*}

In the right-hand panel of Fig.~\ref{fig:GroundPartScaling} we show the square root of the in-ice radiation energy as a function of the total ground particle energy. Each data point corresponds to a given simulation at $10^{\rm 17.5}\, \rm eV$, but with different zenith angle each, resulting in different values of $E_{\rm ground}^{\rm part}$, since the $X_{\rm max}$ distance to the ground increases with the shower zenith angle. The in-ice radiation energy exhibits a clear linear dependence on the energy carried by the shower particles reaching the ice surface within a meter. A linear fit  following
\begin{equation}
    \sqrt{E_{\rm ice}^{\rm rad}} = a\, (E_{\rm part}^{\rm ground})^b \ ,
\end{equation}
yields $a=2.4\times10^{-8}\pm 1\times 10^{-9}$ and $b=0.945\pm 0.06$ which confirms the expected scaling of the in-ice radiation energy with the ground particle energy.

The dependencies of the in-air and in-ice radiation energy evidenced in Fig.~\ref{fig:EnergyScaling}  and Fig.~\ref{fig:GroundPartScaling} are crucial. They allow FAERIE showers simulated at a limited set of energies to be rescaled to other energies, thereby providing large datasets while drastically reducing the computation time. Furthermore, these results should also be valuable for energy reconstruction of the primary particle.


\subsection{Scaling with the zenith angle}~\label{sec:RadE_theta_scaling}

We also studied the dependency of the in-air and in-ice radiation energies with the shower zenith angle. In the left-hand panel of Fig.~\ref{fig:RadEvszen} we show the radiation energy associated to each channel of both the in-air and in-ice emissions, as a function of the shower zenith angle, for simulations with  primary energy $E=10^{\rm 17.5}\, \rm eV$. We observe two opposite behaviors between the in-air and the in-ice component. For all channels, the in-air radiation energy increases with the increasing zenith angle, while the in-ice radiation energy decreases. The trend for the in-ice component is related to the fact that for inclined showers the in-air $X_{\rm max}$ to ice surface distance is larger, which results in a lower ground particle energy and a lower radiation energy as discussed in Section~\ref{sec:RadEscaling}. Because of this dependency, the radio emission from the in-ice component becomes barely detectable for showers with a zenith angle of $50^{\circ}$ (except for the most energetic events) and even weaker for more inclined showers. We note that error bars related to shower to shower fluctuations are larger for the in-ice component than the in-air one, since the in-ice emission varies more rapidly when the air shower maximum fluctuates due to its exponential dependency  with the ground-$X_{\rm max}$ distance evidenced in Section~\ref{sec:RadEscaling}.

\begin{figure*}[t]
    \centering

    \begin{minipage}{0.45\textwidth}
        \centering
        \includegraphics[width=\linewidth]{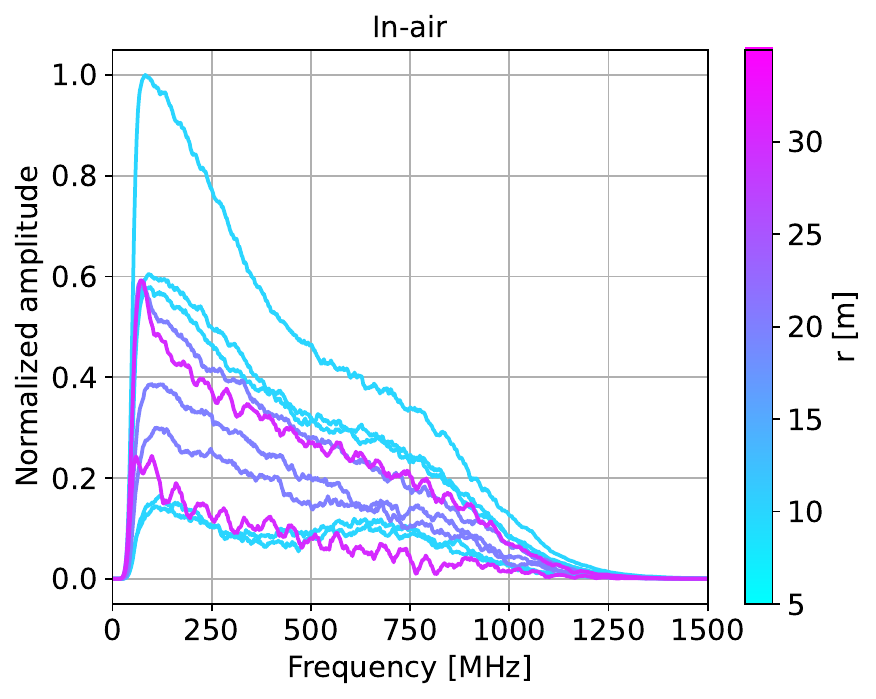}
    \end{minipage}
    \hfill
    \begin{minipage}{0.45\textwidth}
        \centering
        \includegraphics[width=\linewidth]{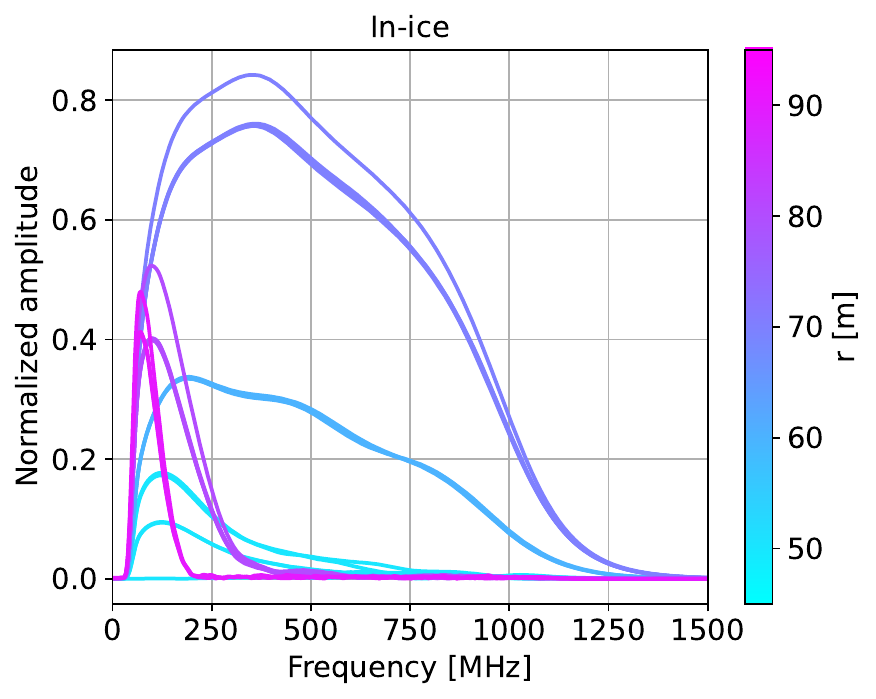}
    \end{minipage}

    \vspace{0.4cm}

    \begin{minipage}{0.48\textwidth}
        \centering
        \includegraphics[width=\linewidth]{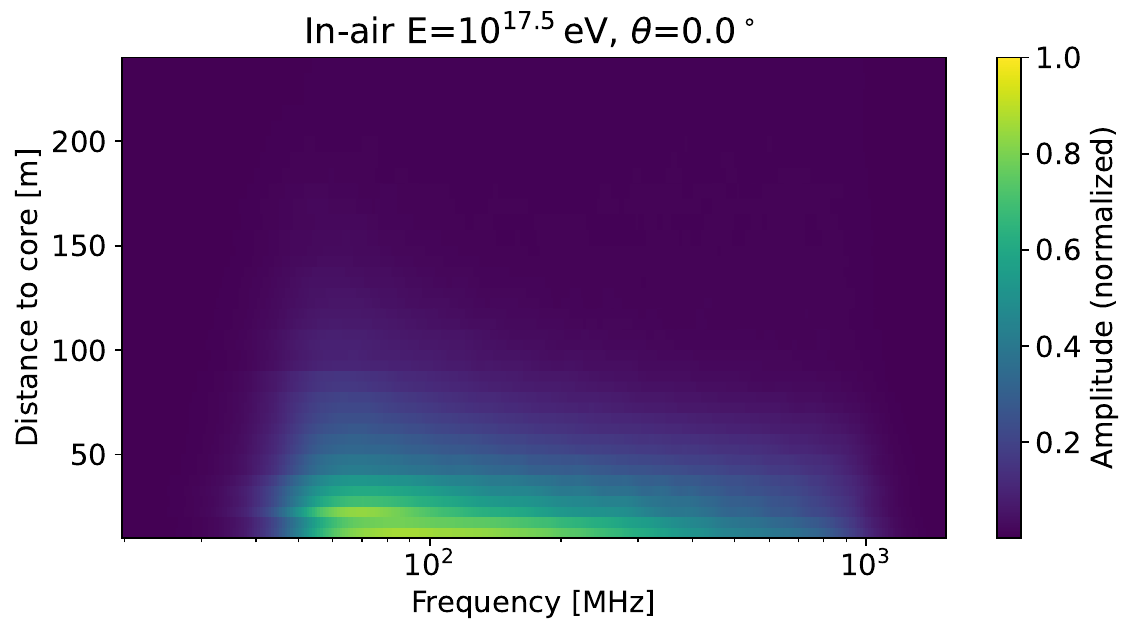}
    \end{minipage}
    \hfill
    \begin{minipage}{0.48\textwidth}
        \centering
        \includegraphics[width=\linewidth]{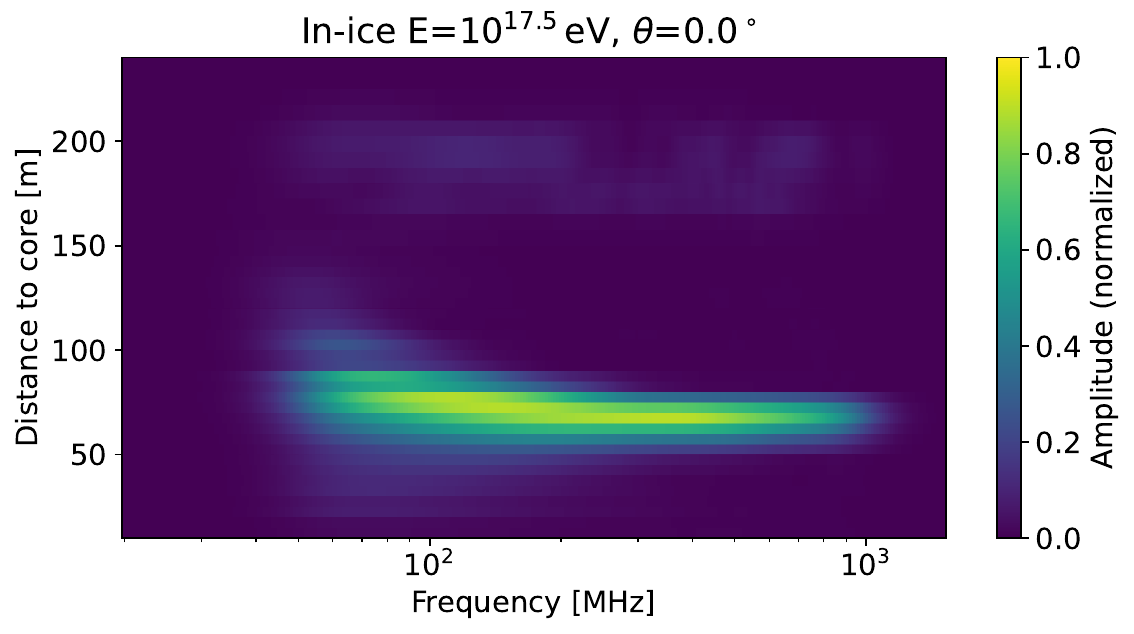}
    \end{minipage}

    \caption{({\it Top}) Frequency-domain amplitude spectra of several antennas for the in-air ({\it left}) and in-ice ({\it right}) emissions, observed at a depth of 100 meters, for a vertical proton shower with primary energy $E=10^{17.5\, \rm eV}$. Antennas were placed at various radial distances (indicated by colors) from the shower core $(r=0\, \rm m)$. The range of radial distances was chosen to cover the region around the Cherenkov cone of each emission component. ({\it Bottom}) 2D heat maps of the in-air and in-ice frequency-domain amplitude spectra, for the same vertical proton shower. The color codes for the amplitude of each frequency, while the $y$-axis indicates the distance of the antennas from the shower core.}
    \label{fig:freqSpectra}
\end{figure*}

For the in-air component, an overall increase of the radiation energy is expected for two main reasons. First, inclined showers develop at higher altitudes (Fig.~\ref{fig:XmaxCharac}), where the lower air density enhances the transverse-current emission (discussion Section~\ref{sec:LDF}). Moreover, the geomagnetic emission scales with $\sin{\alpha}$, with $\alpha$ the so-called geomagnetic angle between the shower and the local magnetic field directions. In Greenland’s near-vertical magnetic field this angle grows with zenith, further enhancing the radiation energy. Additionally, channel-dependent effects are expected as the geomagnetic emission is linearly polarized along the $-\mathbf{v} \times \mathbf{B}$ direction. This direction initially corresponds to the $y$-axis for a vertical shower, but gets projected into the $x$ and $z$ directions when the shower gets inclined, therefore enhancing the signal in these channels.

In the right-hand panel of Fig.~\ref{fig:RadEvszen}, we show the ratio between the in-air and in-ice radiation energies as a function of the shower zenith angle for the three primary energies of our library. It is clear that vertical showers are dominated by the in-ice emission, whereas for more inclined showers $\theta \gtrsim 20^{\circ}$ the in-air component becomes dominant. Increasing the primary energy shifts the air/ice ratio to slightly lower values, as the two contributions scale differently (Section~\ref{sec:RadEscaling}), but the overall trend remains unchanged.  We note that this behavior represents an average feature of the emission. At the scale of individual antennas, the in-air/in-ice ratio may differ depending on the antenna position, but the trend observed in Fig.~\ref{fig:RadEvszen} remains a strong indicator of which component dominates in a given region of the parameter space and of the typical radio signatures (in-air or in-ice like) that should be targeted when reconstructing events. For example, neutrino showers are  expected to arrive at the detector with rather horizontal trajectories, with zenith angles $\theta \gtrsim40^{\circ}$~\cite{RNOG2021JInst..16P3025A}, since they need to travel long enough in the ice to cascade in the medium. Yet, in this zenith angle range the typical cosmic ray signals should be in-air like, i.e., dominated by the geomagnetic emission, hence with quite distinctive features than neutrino signals, which would be valuable for cosmic ray/neutrino discrimination. We note, however, that refraction of radio waves in the ice bends the signal trajectories, so that the emission may reach the detector with an apparent zenith angle different from the geometric shower direction, which would soften  the discriminating power of an observable based on zenith angle to identify neutrinos.

\section{Radio signal observables}~\label{sec:Observables}
Once we know which emission component of cosmic ray emission should be dominant, we can evaluate the characteristic signatures  of each emission, that could be targeted to identify cosmic rays or discriminate them from neutrino primaries when analyzing experimental data.

\subsection{Frequency signatures}~\label{sec:frequency}

In-ice detectors such as ARA and RNO-G operate over a broad frequency range, typically from $\sim 100$ to 800 MHz. This raises the possibility of using the frequency content of the radio signal to identify cosmic-ray signatures. The in-air and in-ice emissions from cosmic-ray showers are expected to exhibit distinctive frequency features: the in-air cascade has a shower front thickness of the order of a few meters, limiting coherent emission to relatively low frequencies. In contrast, for the in-ice cascade, coherence is limited by the lateral particle distribution, which extends over only a few centimeters, allowing the signal to remain coherent up to much higher frequencies.

In Fig.~\ref{fig:freqSpectra}, we show frequency signatures of the in-air (left-hand panel) and in-ice (right-hand panel) emissions from FAERIE simulations, for a vertical proton shower with an energy $E=10^{\rm 17.5}\, \rm eV$, observed at a depth of 100 meters. The top panels show the frequency-domain amplitude spectra of a few antennas located around the Cherenkov ring of each emission. The plot  highlights that the in-air emission has a rather sharp amplitude spectrum that peaks at low-frequency below $100\, \rm MHz$. On the other hand, the in-ice emission shows a broad amplitude spectrum which peaks at much higher frequencies around $\sim 400 \, \rm MHz$ for the antennas located on the Cherenkov ring $r_{\rm cer}^{\rm ice}\sim70\, \rm m$. These features could already enable cosmic-ray identification or emission-mechanism discrimination at the single-antenna level, using for example the ratio between the high to low frequency power, as was reported, for example, by the ARA collaboration~\cite{ARAcr2025arXiv251021104A}.

The bottom panels of Fig.~\ref{fig:freqSpectra} show 2D heatmaps of the amplitude spectra as a function of the antenna distance from the shower core. These plots reveal that the in-air emission (left-hand panel) peaks at low frequencies and at small core distances, whereas for the in-ice emission (right-hand panel) the highest amplitudes occur at much higher frequencies and at larger distances, around 70 meters away from the core. The figure also illustrates the strong spatial variability of the in-ice signal: the emission is sharply confined around the Cherenkov cone, and shifts to lower frequencies both before and beyond the  cone. This indicates that in the case where several antennas are triggered, the spatial variation of the frequency content could yield relevant information to identify cosmic rays. The map also exhibits broadband, low-amplitude frequency content around radial distances of about 200 m. This corresponds to the location of the double ring in the same simulation (see Fig.~\ref{fig:footprints}). In the time domain, this feature appears as an interference pattern with alternating peaks and dips. Its origin is currently unclear, and will be investigated further. Finally, we note that though the results are shown for only one shower, similar behavior is observed for showers with different primary energy and zenith angles.

\begin{figure}[ht]  
    \centering
    \includegraphics[width=\linewidth]{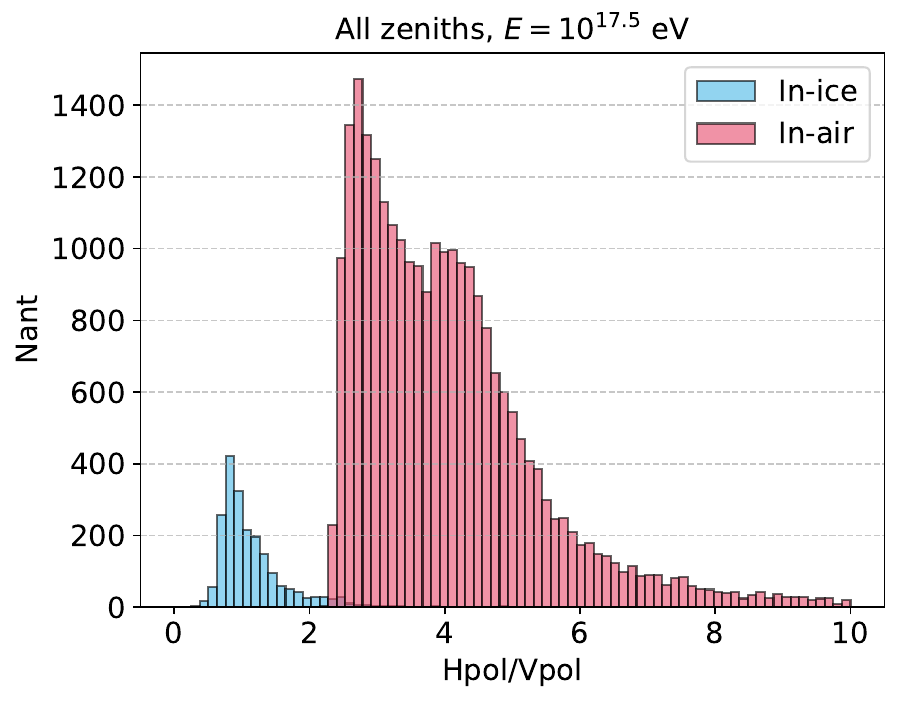}
    \caption{Distribution of the Hpol/Vpol ratio of the in-air (red) and in-ice (blue) components, observed at a depth of 100 meters, from FAERIE simulations for showers with primary energy $E=10^{17.5}\, \rm eV$, and zenith angle between $0^{\circ}$ and $50^{\circ}$. The Hpol signal is defined as $E^{\rm Hpol} = \sqrt{E_x^2 + E_y^2}$, the Vpol signal is defined as $E^{\rm Vpol} = E_z$}.
    \label{fig:PolarDistrib}
\end{figure}

\begin{figure*}[tb]
\includegraphics[width=0.49\linewidth]{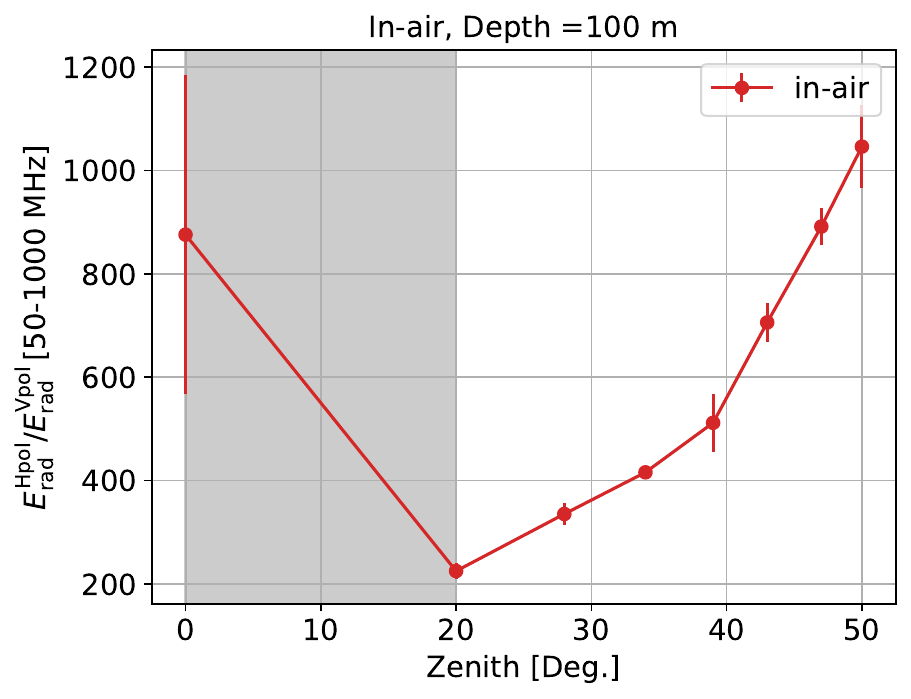}\hfill
\includegraphics[width=0.49\linewidth]{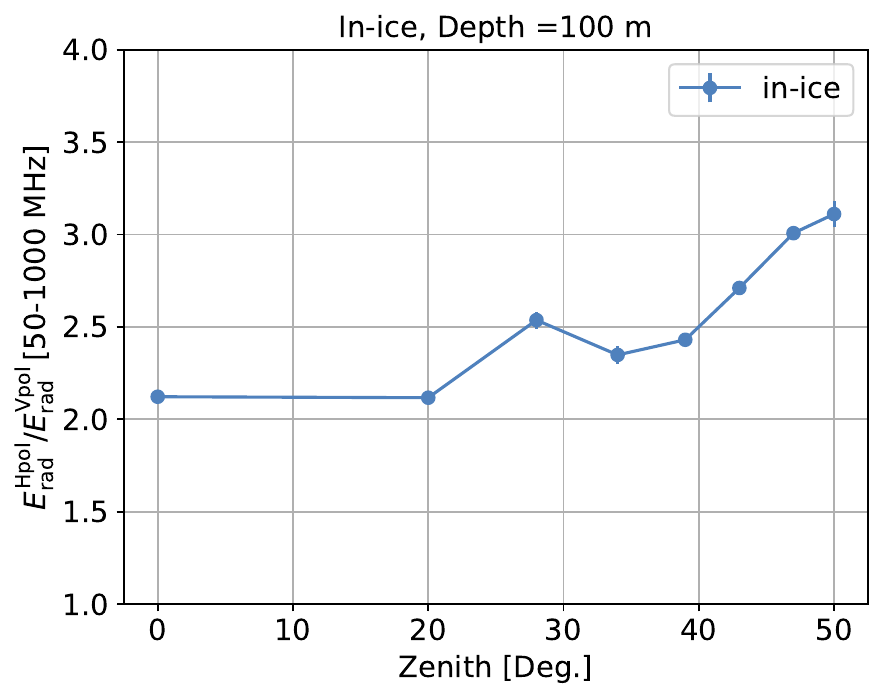}
\caption{Ratio of the radiation energy measured in the Hpol and Vpol antennas for the in-air ({\it left}) and in-ice ({\it right}) components, as a function of the shower zenith angle. The results are averaged over the three primary energies of our library, the dots correspond to the mean and the error bars to the RMS of the distribution. The gray shaded region in the left-hand panel indicates the zenith-angle interval $0^{\circ}$–$25^{\circ}$, where the in-air $X_{\rm max}$ can lie below the ice surface, thereby modifying the expected polarization ratio.}\label{fig:PolarRadE}
\end{figure*}


\subsection{Polarization signatures}~\label{sec:Polarization}

Using our simulation library, we also investigated the polarization of the radio signal of each emission. Detectors like ARA and RNO-G typically combine Hpol antennas, that measure the horizontal polarization  of the radio signal, with Vpol antennas, that measure the vertical polarization of the emission~\cite{RNOinst2025JInst..20P4015A}. The signal in the Vpols can be estimated using the $z$-channel of the electric field $E^{\rm Vpol}=E_z$, while the signal in the Hpols can be estimated using the quadratic sum of the signal in the $x$ and $y$ channels $E^{\rm Hpol}=\sqrt{E_x^2+ E_y^2}$. Using these definitions, we can then investigate the polarization of the radio signal using the Hpol/Vpol ratio. In practice, in-ice neutrino detectors use antenna that are typically sensitive to the electric field components orthogonal to the signal arrival direction. Estimating these components would require ray tracing the radio emission to each antenna. For simplicity, we therefore approximate the Hpol and Vpol contributions directly from the Cartesian field components.

In Fig.~\ref{fig:PolarDistrib}, we show the distributions of the Hpol/Vpol peak amplitude ratio at the antenna level, at a depth of 100 m, for simulations with a primary energy of  $E=10^{17.5} \, \rm eV$, covering all zenith angles in our library. The in-air distribution contains more antennas than the in-ice one, as the in-air footprint is larger and therefore a greater number of antennas have a well-defined Hpol/Vpol ratio. The figure evidences that the in-air emission shows stronger Hpol signals and hence much higher Hpol/Vpol ratios than the in-ice component. This is expected, as the geomagnetic emission, which is the dominant component for the in-air mechanism, mainly projects in the horizontal plane, as shown in the left-hand panel of Fig.~\ref{fig:RadEvszen}. Both in-air and in-ice distributions can clearly be resolved, implying that already at the antenna level, the polarization of the radio signal could help discriminate between the in-air and in-ice emissions from cosmic ray showers. Furthermore, since the neutrino emission is typically Askaryan-like, this also implies that the polarization could help discriminate between the in-air cosmic ray emission and the in-ice Askaryan emission from neutrino showers. We note, however, that in practice we will not directly measure the radio signal polarization, but rather the signal voltage which corresponds to the projection of the electric field along the antenna effective length. This will of course deteriorate the polarization resolution. Still, some of the features observed at the electric field should remain and could still be visible in the voltage.

\begin{figure*}[tb]
\includegraphics[width=0.51\linewidth]{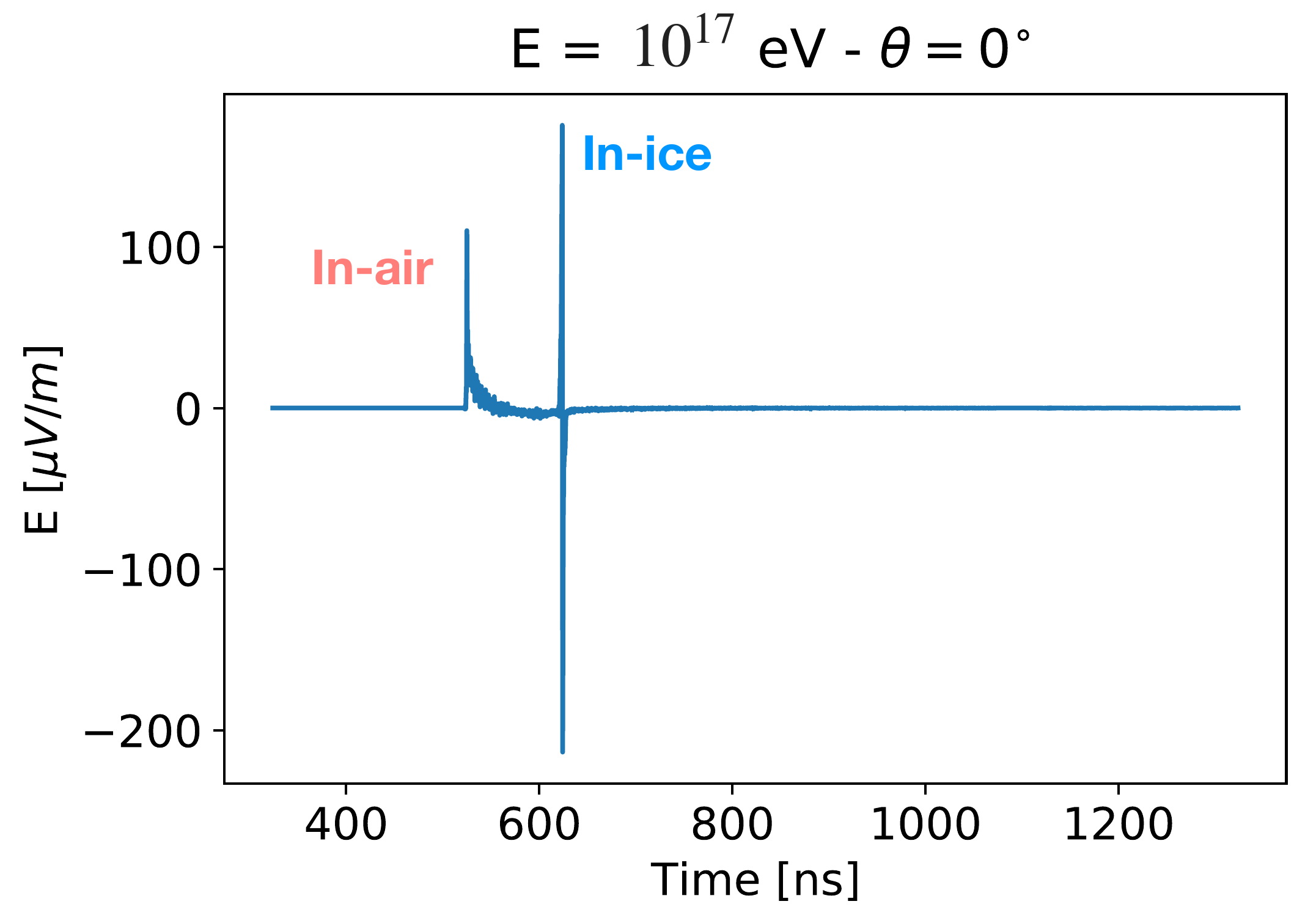}\hfill
\includegraphics[width=0.47\linewidth]{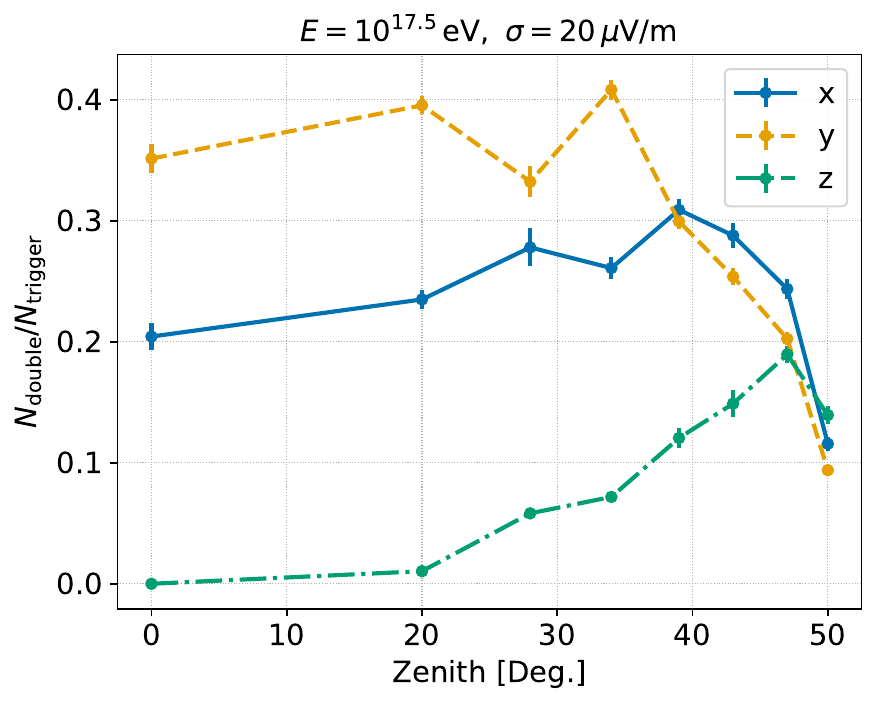}
\caption{({\it Left}) Typical double pulse signal measured at a given antenna at a depth of 100 meters, for the East-West component of the electric field, from FAERIE simulations. ({\it Right}) Double pulse rate per channel, observed at a depth of 100 meters, as a function of the shower zenith angle, for primary particles with energy $E=10^{17.5}\, \rm eV$. The double pulse rate is given by the number of antennas seeing a double-pulse events divided by the number of triggered antennas. The error bars represent the estimated uncertainties related to shower-to-shower fluctuations.}\label{fig:DoublePulseChannels}
\end{figure*}

\begin{figure}[tb]  
    \centering
    \includegraphics[width=\linewidth]{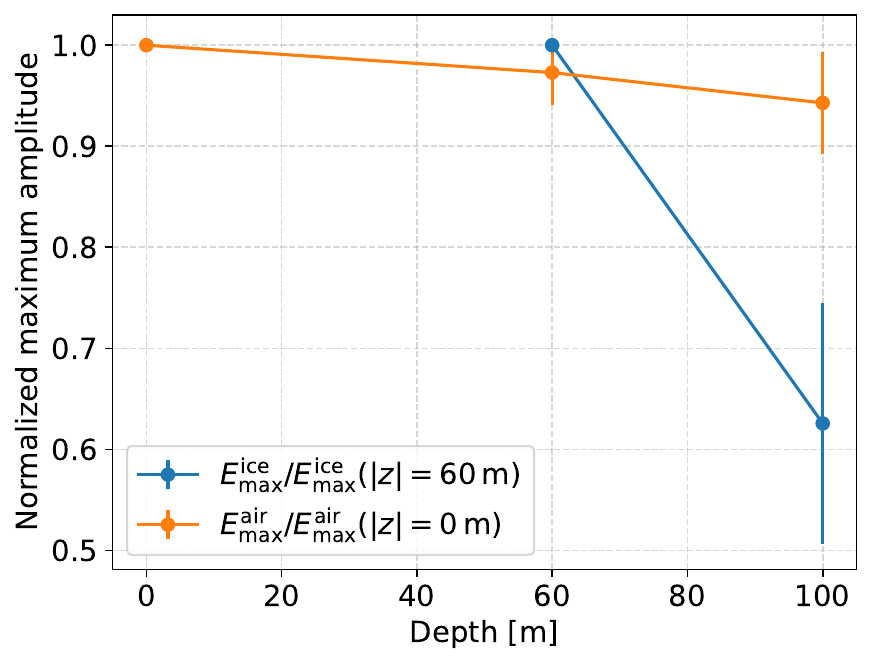}
    \caption{In-air (orange) and in-ice (blue) maximum signal amplitude versus antenna depth. The results are averaged over the 8 zenith angles of our library and the error bars show the standard deviations of the distributions. The values are normalized by their respective maximum signal amplitude at the ice surface (for the in-air emission), or at a depth of 60 meters (for the in-ice emission).}
    \label{fig:SignalVsDepth}
\end{figure}

\begin{figure*}[tb]
\includegraphics[width=0.49\linewidth]{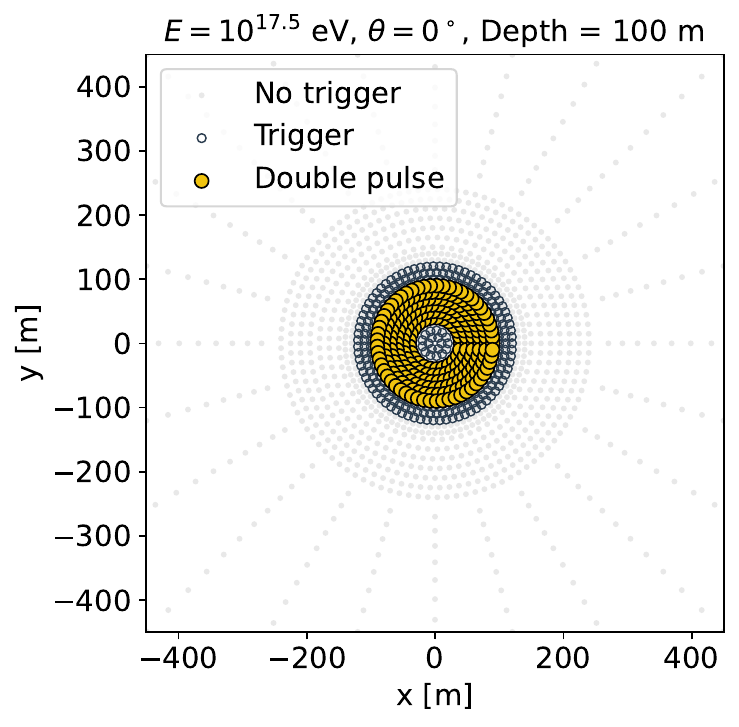}\hfill
\includegraphics[width=0.49\linewidth]{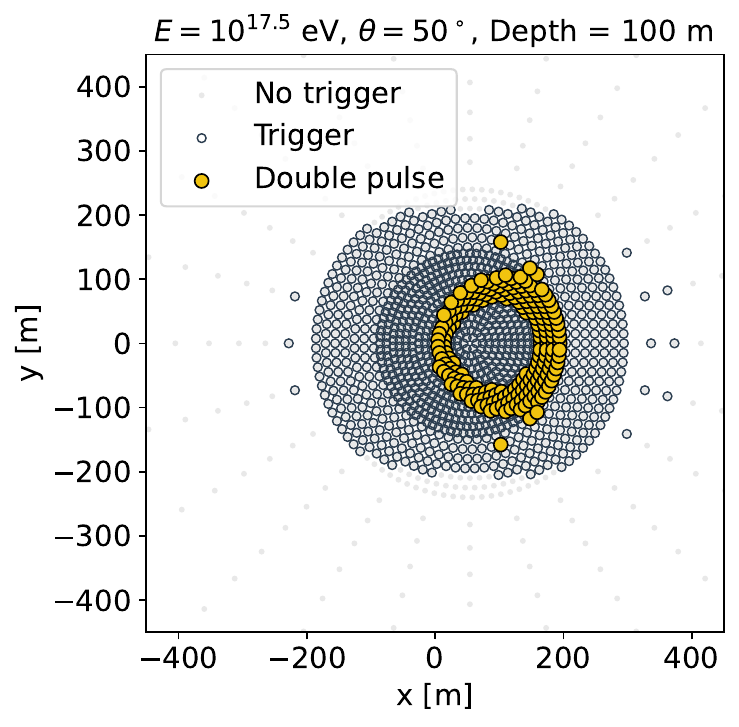}
\caption{{Spatial distribution of double-pulse signals for vertical ($\theta = 0^{\circ}$) and inclined ($\theta = 50^{\circ}$) showers with a primary energy of $E = 10^{17.5},\mathrm{eV}$, observed at a depth of 100 m. Yellow markers indicate antennas detecting a double pulse, dark blue markers correspond to antennas that trigger but record only a single pulse, and gray markers represent antennas that do not trigger.}}\label{fig:DoublePulseDensity}
\end{figure*}

We can also evaluate the total radiation energy measured by the Hpol and Vpol antennas by substituting $E_{\rm x,y,z}$ in Eqs.~\eqref{eq:fluence} and \eqref{eq:RadE}. This  yields $E_{\rm rad}^{\rm Hpol}= \sqrt{(E_x^{\rm rad})^2+(E_y^{\rm rad})^2}$ and  $E_{\rm rad}^{\rm Hpol}=E_z^{\rm rad}$.
In Fig.~\ref{fig:PolarRadE}, we show the ratio of Hpol and Vpol radiation energies $E_{\rm rad}^{\rm Hpol}/E_{\rm rad}^{\rm Vpol}$,  as a function of the shower zenith angle, for both the in-air (left-hand panel) and in-ice emissions (right-hand panel). The results are averaged over the three primary energies of our library, the dots indicate the average results while the error bars show the standard deviations of the distributions. As the radiation energy corresponds to an average result over many antennas we can see that the difference in the Hpol/Vpol ratio becomes even more relevant on these plots with more than 2 orders of magnitude difference, independently of the shower zenith angle. For the in-air emission, the vertical case stands out, as it corresponds to a case where the in-air $X_{\rm max}$ can be below the ice for the most energetic showers (Fig.~\ref{fig:XmaxCharac}), which also explains the large error bars, when averaging over several energies. For more inclined showers we can observe a continuous increase of the Hpol/Vpol ratio with increasing zenith angle. This trend arises because the geomagnetic emission strengthens for inclined showers—due to both the lower density at $X_{\rm max}$ and the larger geomagnetic angle (see Section~\ref{sec:RadE_theta_scaling})—while the Askaryan contribution weakens as the reduced air density at $X_{\rm max}$ suppresses the charge-excess emission~\cite{GlaserRadE2016JCAP...09..024G, ChichePolar2022APh...13902696C}. For the in-ice emission, the radial polarization imposes similar amplitudes in each channel as shown in Fig.~\ref{fig:RadEvszen}. This yields  a roughly constant ratio around 2, since the radiation energy scales with the squared electric field. For more inclined showers, a slight increase is observed, which we associate with ray bending effects that become more prominent for more horizontal showers and that will shift the polarization. These results highlight that in the case where an event is seen by several antennas, averaging the signals over the antennas should make the polarization signature even more significant.

\subsection{Signal amplitude dependence on antenna depth}

As mentioned in Section~\ref{sec:LDF}, the signal amplitude of the in-ice emission is expected to vary more strongly with antenna depth than that of the in-air emission, since its source is much closer to the antennas. To highlight this behavior, Fig.~\ref{fig:SignalVsDepth} shows the evolution of the in-air and in-ice signal amplitudes as a function of antenna depth. For each depth, we select the antenna with the maximum amplitude in order to compare signals originating from antennas located near the Cherenkov cone and therefore roughly at the same position within the radio footprint. The results are averaged over the eight zenith angles of our library, and the error bars represent the standard deviation of the distributions.

The signal from the in-air emission remains approximately constant with antenna depth. Only a slight decrease in amplitude is observed for the deepest antennas, a trend mainly driven by the most vertical showers, whose in-air $X_{\rm max}$ lies closer to the ice surface. In contrast, the in-ice emission shows a strong decrease in signal amplitude with increasing antenna depth. Larger fluctuations are observed because, for inclined showers, the in-ice emission is significantly bent by refraction in the ice. Nevertheless, the average in-ice signal amplitude at a depth of 100 m corresponds to about 60\% of that at 60 m depth, consistent with the expected behavior for a signal emitted near the ice surface and propagating approximately vertically downward. The strong dependence of the signal amplitude on antenna depth therefore provides a useful handle to identify signals emerging from the noise and to reject far-field noise sources.


\subsection{Double pulses}~\label{sec:DoublePulse}

In the case where both emissions overlap, the in-air and in-ice signals from cosmic ray showers can sometimes reach the same antenna and give rise to a typical double pulse signal~\cite{KDV2016APh....74...96D}.  We illustrate such a signal by showing the East-West component of the electric field of a given antenna in the left-hand panel of Fig.~\ref{fig:DoublePulseChannels}.

The in-air pulse generally arrives before the in-ice one. To first order, the in-air emission propagates approximately along a direct path to the observer. The in-ice signal, in contrast, is generated after the cascade develops in the ice and is emitted at the Cherenkov angle $\theta^{\rm ice}_c \sim 42^{\circ}$ at the in-ice shower maximum, resulting in a longer path length and hence a delay.

Using FAERIE simulations, we can investigate the occurrence of double pulse events. Setting the background level at $\sigma =20\, \rm \mu V/m$, we assume that a double pulse is observed at a given antenna if (1) there is at least one pulse with a peak amplitude in the norm of the electric field above $5\sigma$, (2) there is a second pulse with a peak amplitude above $3\sigma$ and (3) the two pulses are separated by at least $50\, \rm ns$. These conditions ensure that the antenna ``triggers'', that second low-amplitude pulse can be distinguished from noise, and that the in-air and in-ice pulses remain temporally resolvable, even in the presence of additional timing uncertainties introduced by the detector response. In Fig.~\ref{fig:DoublePulseDensity} we show maps of the double pulse spatial distribution at a depth of 100 meters, for a vertical shower ($\theta =0^{\circ}$) and an inclined shower ($\theta =50^{\circ}$) with primary energy $E=10^{17.5}\, \rm eV$. By comparing with the fluence footprints in Fig.~\ref{fig:footprints}, we see that the double-pulse distributions generally follow the in-ice footprints. For the vertical shower, this is mainly due to the strong spatial variability of the in-ice emission, whose amplitude peaks at the Cherenkov cone and drops rapidly away from it. For the inclined shower, the in-air footprint encompasses the in-ice one, whose amplitude is subdominant. A double pulse therefore appears only when the in-ice emission exceeds the thresholds defined above. Increasing the trigger thresholds or decreasing the primary particle energy yields globally similar results, with a reduced number of double pulse events. Typically, at lower energy, the map for the vertical shower exhibits a deficit of double pulses where the geomagnetic and charge-excess mechanisms interfere destructively, due to the in-air emission becoming subdominant.

Finally, we investigate the rate of double-pulse events observed in each channel. For each simulation, we estimate the number of antennas exhibiting a double pulse,  $N_{\rm double}$, using the criteria defined above. Since this analysis is performed at the individual channel level rather than on the electric-field norm, the threshold conditions are divided by a factor $\sqrt{3}$. Likewise, for each channel we determine the total number of triggered antennas, $N_{\rm triggered}$, defined as those with a peak electric-field amplitude exceeding $5\sigma/\sqrt{3}$. The double pulse rate is then estimated from the ratio of these two quantities as $N_{\rm double}/N_{\rm triggered}$. In the right-hand panel of Fig.~\ref{fig:DoublePulseChannels}, we present the double-pulse rate per channel at a depth of 100 m as a function of the shower zenith angle for simulations with a primary energy of $E=10^{17.5}\, \rm eV$. A significant fraction of triggered events exhibit a double pulse, reaching up to $\sim 40\%$ in the $y$-channel. The dependence on zenith angle can be understood as follows. For nearly vertical showers, the in-air component is subdominant (see Fig.~\ref{fig:RadEvszen}). As the shower inclination increases, the in-air radiation energy grows in all channels and its spatial extent broadens, leading to a corresponding increase in the double-pulse rate. For the most inclined showers, however, the in-ice emission becomes subdominant and too weak to produce a clearly distinguishable pulse, causing the double-pulse rate to decrease. Consequently, the highest rate of double pulse events is observed at the intermediate zenith angle of $\theta\sim 34^{\circ}$. The figure also shows that double pulses are far more likely to be observed in the Hpol antennas (high rate in the $x$ and $y$ channels) than in the Vpol antennas. Hence, comparing the rate of double pulses in these two types of antennas could help identify cosmic ray signals or reconstruct their zenith angle. Similar results are observed when using the antenna layer at a depth of 60 meters. Decreasing the primary energy or increasing the trigger thresholds can also significantly decrease the observed double pulse rate while roughly preserving the global trend. For example, the maximum rate of the double pulse events  falls to $\sim 20\%$ for showers with primary energy $E=10^{16.5}\, \rm eV$, and no double pulse is observed in the $z$-channel for showers with primary energies below or equal to $E=10^{17}\, \rm eV$.

\section{Discussion}~\label{sec:Discussion}

The radio emission from cosmic-ray particle showers exhibits distinctive features that can guide future analyses and help their identification with in-ice radio detectors. Furthermore, the differences between the in-air and in-ice components will be valuable to discriminate between both, but also to discriminate the in-air cosmic ray emission from neutrinos, since neutrinos signatures should typically be Askaryan-like. We note that these results were obtained at the electric field level, using raw outputs of FAERIE simulations. Hence, to evaluate  the true background rejection efficiency and discrimination power that could be reached, one would need to incorporate the detector response chain and the ambient noise background.  Indeed, in experimental conditions, signatures such as the radio signal polarization, frequency and amplitude distributions will be modified by the antenna’s effective length, total system gain and bandwidth... Still, the main features discussed in this article should remain and for example a recent study on the first observation of in-ice cosmic ray signals, performed by ARA typically used some of the radio signal observables mentioned in this study~\cite{ARAcr2025arXiv251021104A} (polarization and frequency spectrum).

Identifying cosmic rays  with in-ice radio detectors will be challenging, and  towards that goal, experiments could rely on the arrival direction of the signals. Indeed, cosmic rays interact first in the atmosphere hence detectors like RNO-G use surface antennas to veto cosmic rays and discriminate them from neutrino primaries. Reconstructing the true arrival direction of the radio emission should also be valuable, as we have shown that in-air and in-ice cosmic ray signatures both have a strong dependence on the shower zenith angle. Particularly, the fact that the in-air component dominates for inclined showers (Section~\ref{fig:RadEvszen}) should drastically ease the discrimination with neutrino emission. The rapid variation of the in-ice signal with the antenna depth mentioned in Section~\ref{sec:LDF} also provides a valuable way to remove any far above surface background source. Double pulses will be the most qualitative detectable events, allowing us to observe characteristics of the in-air and in-ice emissions, at the single antenna level, after they went through the same detection chain.  At the highest energies ($E\geq 10^{17.5}\, \rm eV$), these events will represent a significant fraction of triggered events  (Section~\ref{sec:DoublePulse}). Their observation, with the first pulse matching the characteristics of the in-air emission and the second pulse matching the in-ice emission would be a smoking gun of cosmic ray emission. The scaling relations evidenced in Sections~\ref{sec:RadEscaling} and~\ref{sec:RadE_theta_scaling} could therefore allow us to build huge datasets of simulations and to reconstruct cosmic rays using template matching methods. 

Eventually, we further note that our simulations used Greenland magnetic field has a similar amplitude and direction to the magnetic field at the South Pole, a location which can also host in-ice radio detectors. The main difference between the two magnetic fields is the polarity of the $z$-component (downwards in Greenland and upwards in South Pole). Hence our results could also be partly relevant for South Pole experiments such as the Askaryan Radio Array and the planned IceCube-Gen2 radio detector.


\section{Conclusion}~\label{sec:Conclusion}

Particle cascades induced by high energy cosmic rays will emit in-air and in-ice radio emission that could be detected with in-ice neutrino detectors such as ARA, RNO-G, and IceCube-Gen2 radio. Using FAERIE Monte-Carlo simulations, we investigated radio signatures of cosmic ray particle cascades, based on the signal amplitude, radiation energy, frequency, polarization, and timing. Our results show that the in-ice component from cosmic ray emission should be dominant for vertical showers, while the in-air component dominates for inclined showers, with zenith angle $\theta \gtrsim 20^{\circ}$. We further showed that the in-air and in-ice component have different signatures: The emission from the in-air component peaks at low frequency (below $100\, \rm MHz)$, close to the shower core and has a dominant Hpol polarization. On the other hand, the emission from the in-ice component peaks at higher frequencies (around $400\, \rm MHz$), further away from the shower core and shows a radial polarization. Finally, we investigated the spatial distribution of double pulse events. We found that double pulse signals are more likely to be observed by horizontally polarized channels, than vertically polarized ones, and that for primary energies $E>10^{17.5}\, \rm eV$, their rate could represent up to $\sim 40\%$ of triggered events in the $y$-channel of the electric field. 

These results provide a consistent physical framework to interpret the in-ice radio emission from cosmic-ray–induced cascades. The distinct spectral, polarization, and spatial features identified in this work offer practical observables for identifying cosmic-ray events, reconstruct their geometry and energy, and calibrate in-ice radio detectors. In addition, they provide key observables to discriminate cosmic rays from neutrino-induced signals in the presence of a dominant cosmic-ray background, thereby strengthening the prospects for ultra-high-energy neutrino detection with future in-ice radio experiments.






\subsection*{Acknowledgments}

S. Toscano and S. Chiche are supported by the  Belgian Funds for Scientific Research (FRS-FNRS). We thank the referees for their thorough read of the article and their insightful feedback.


\bibliographystyle{elsarticle-num} 
\bibliography{biblio}

\end{document}